\documentclass[sigconf,screen,authorversion,nonacm]{acmart}

\AtBeginDocument{%
  \providecommand\BibTeX{{%
    \normalfont B\kern-0.5em{\scshape i\kern-0.25em b}\kern-0.8em\TeX}}}

\setcopyright{acmcopyright}
\copyrightyear{2018}
\acmYear{2018}
\acmDOI{XXXXXXX.XXXXXXX}

\acmConference[Conference acronym 'XX]{Make sure to enter the correct
  conference title from your rights confirmation emai}{June 03--05,
  2018}{Woodstock, NY}
\acmPrice{15.00}
\acmISBN{978-1-4503-XXXX-X/18/06}

\usepackage[inline]{enumitem}
\usepackage{multirow}
\usepackage{booktabs}
\usepackage{tabularx}
\usepackage{subfig}

\begin{document}

\title[Tangi]{Tangi: a Tool to Create Tangible Artifacts for Sharing Insights from 360° Video.}
\author{Wo Meijer}
\email{W.I.M.T.Meijer@tudelft.nl}
\orcid{0000-0002-8369-6394}
\affiliation{%
  \institution{TU Delft}
  \streetaddress{Landbergstraat 15}
  \city{Delft}
  \state{Zuid-Holland}
  \country{Netherlands}
  \postcode{2628 CE}
}
\author{Jacky Bourgeois}
\orcid{0000-0003-1090-5703}
\email{j.bourgeois@tudelft.nl}
\affiliation{%
  \institution{TU Delft}
  \streetaddress{Landbergstraat 15}
  \city{Delft}
  \country{The Netherlands}
  \postcode{2628 CE}
}
\author{Tilman Dingler}
\orcid{0000-0001-6180-7033}
\email{T.Dingler@tudelft.nl}
\affiliation{%
  \institution{TU Delft}
  \streetaddress{Landbergstraat 15}
  \city{Delft}
  \country{The Netherlands}
  \postcode{2628 CE}
}
\author{Gerd Kortuem}
\orcid{0000-0003-3500-0046}
\email{g.w.kortuem@tudelft.nl}
\affiliation{%
  \institution{TU Delft}
  \streetaddress{Landbergstraat 15}
  \city{Delft}
  \country{The Netherlands}
  \postcode{2628 CE}
}

\renewcommand{\shortauthors}{Meijer, et al.}
\begin{abstract}
Designers often engage with video to gain rich, temporal insights about the context of users, collaboratively analyzing it to gather ideas, challenge assumptions, and foster empathy.
To capture the full visual context of users and their situations, designers are adopting 360° video, providing richer, more multi-layered insights.
Unfortunately, the spherical nature of 360° video means designers cannot create tangible video artifacts such as storyboards for collaborative analysis.
To overcome this limitation, we created Tangi, a web-based tool that converts 360° images into tangible 360° video artifacts, that enable designers to embody and share their insights.
Our evaluation with nine experienced designers demonstrates that the artifacts Tangi creates enable tangible interactions found in collaborative workshops and introduce two new capabilities: spatial orientation within 360° environments and linking specific details to the broader 360° context.
Since Tangi is an open-source tool, designers can immediately leverage 360° video in collaborative workshops.
\end{abstract}

\begin{teaserfigure}
 \centering
    \subfloat[Equirectangular projection.]{\label{fig:teaser-normal}\includegraphics[width=0.45\textwidth]{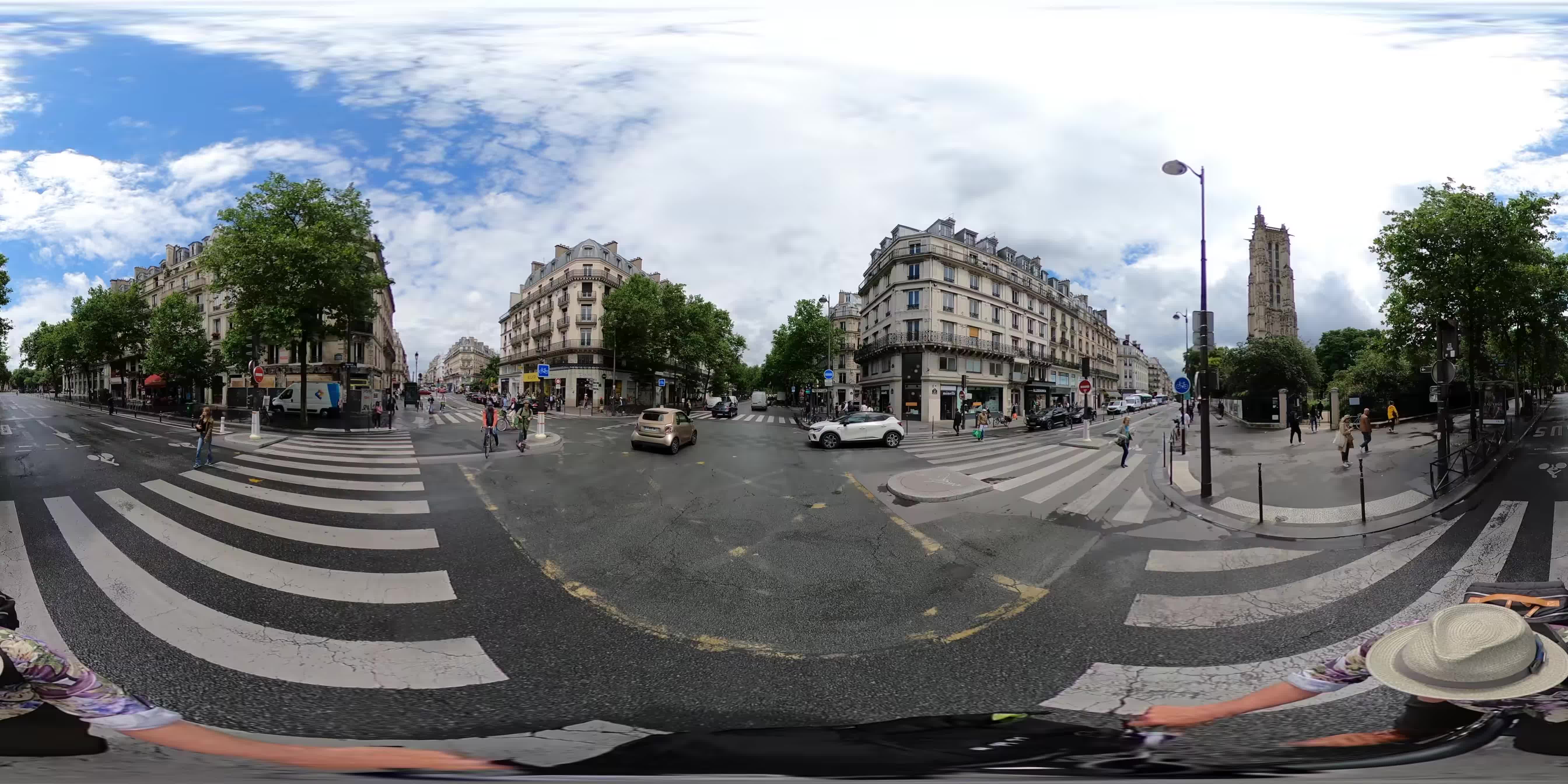}}
    \qquad
   \subfloat[Cuboctahedron artifact created by Tangi.]{\label{fig:teaser-poly}\includegraphics[width=0.45\textwidth]{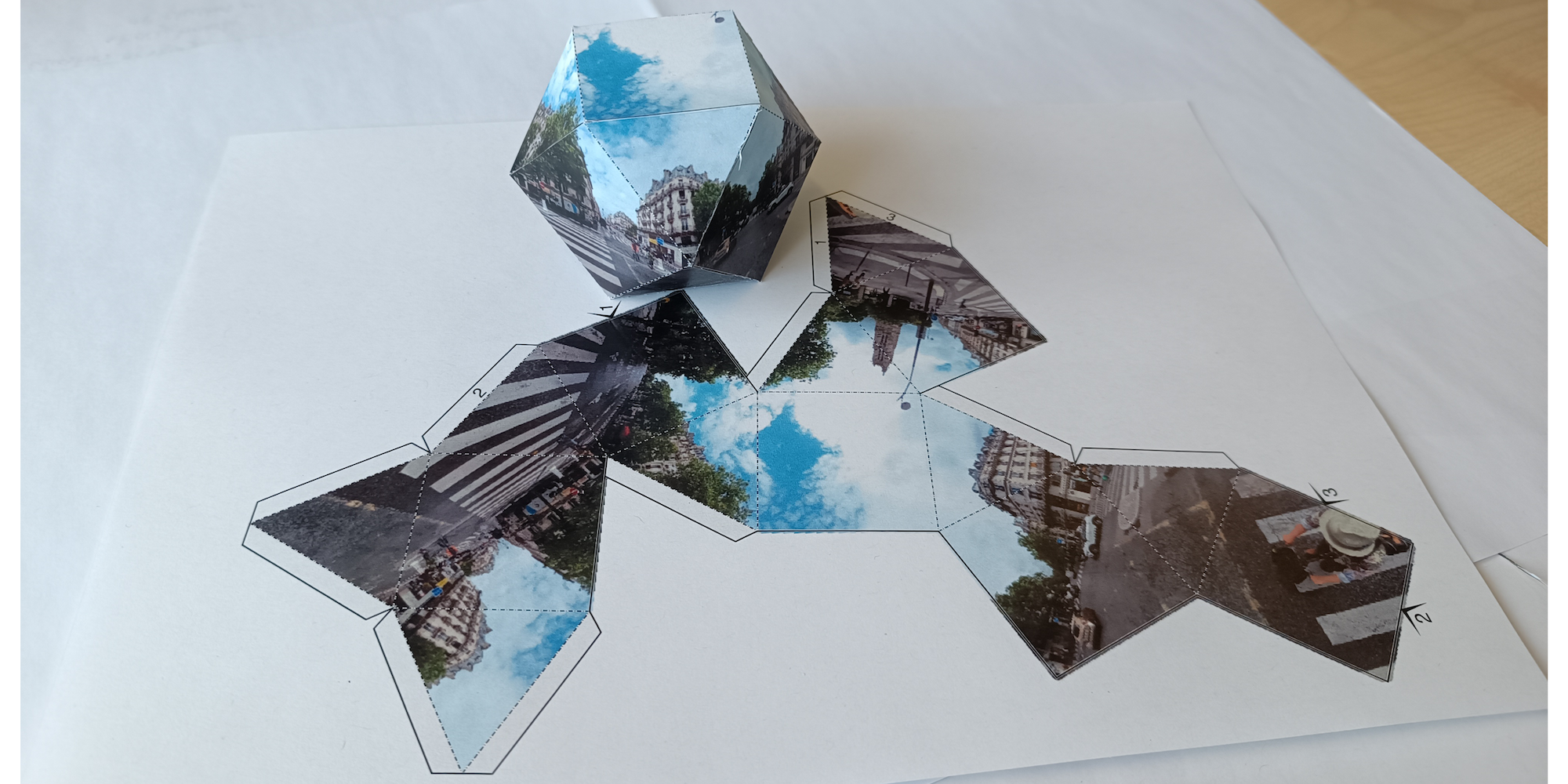}}
  \caption{Printing out frames of 360° video leads to highly distorted artifacts (a) that make collaboration challenging. The artifacts generated by Tangi (b) enable designers to share, collaborate, annotate, and document insights from 360° video with minimal distortion.}
  \Description{The same frame of 360° video shown as a flat image (\ref{fig:teaser-normal}) and as a cuboctahedron artifact (\ref{fig:teaser-poly}) created by Tangi. Video from \protect\href{http://www.velomondial.net/}{Velo Mondial} - \protect\href{https://creativecommons.org/licenses/by/4.0/}{CC-BY}.}
  \label{fig:teaser}
\end{teaserfigure}
\received{20 February 2007}
\received[revised]{12 March 2009}
\received[accepted]{5 June 2009}

\maketitle

\section{Introduction}
To better understand the needs and wants of potential users, designers\footnote{Someone engaged in the processes of (re)designing a product or service, regardless of profession or title.} engage in Contextual Inquiry, gaining insights into the context around the user and the user themselves~\cite{beyer_contextual_1999}. One method for gathering information about a context is the use of video, which enables prolonged and unobtrusive observation of a context~\cite[p.~19]{ylirisku_designing_2007} or observation of contexts that are difficult or dangerous to observe in person -- for example, logging equipment operators~\cite{lamas_analyzing_2019} or emergency medical services~\cite{schlosser_designing_2022}. 
\begin{figure*}[!ht]
 \centering
 \subfloat[A subsection of a 360° video cropped to match the FoV (Field of View) of conventional video.]{\label{fig:conventional-video}\includegraphics[width=0.4\textwidth]{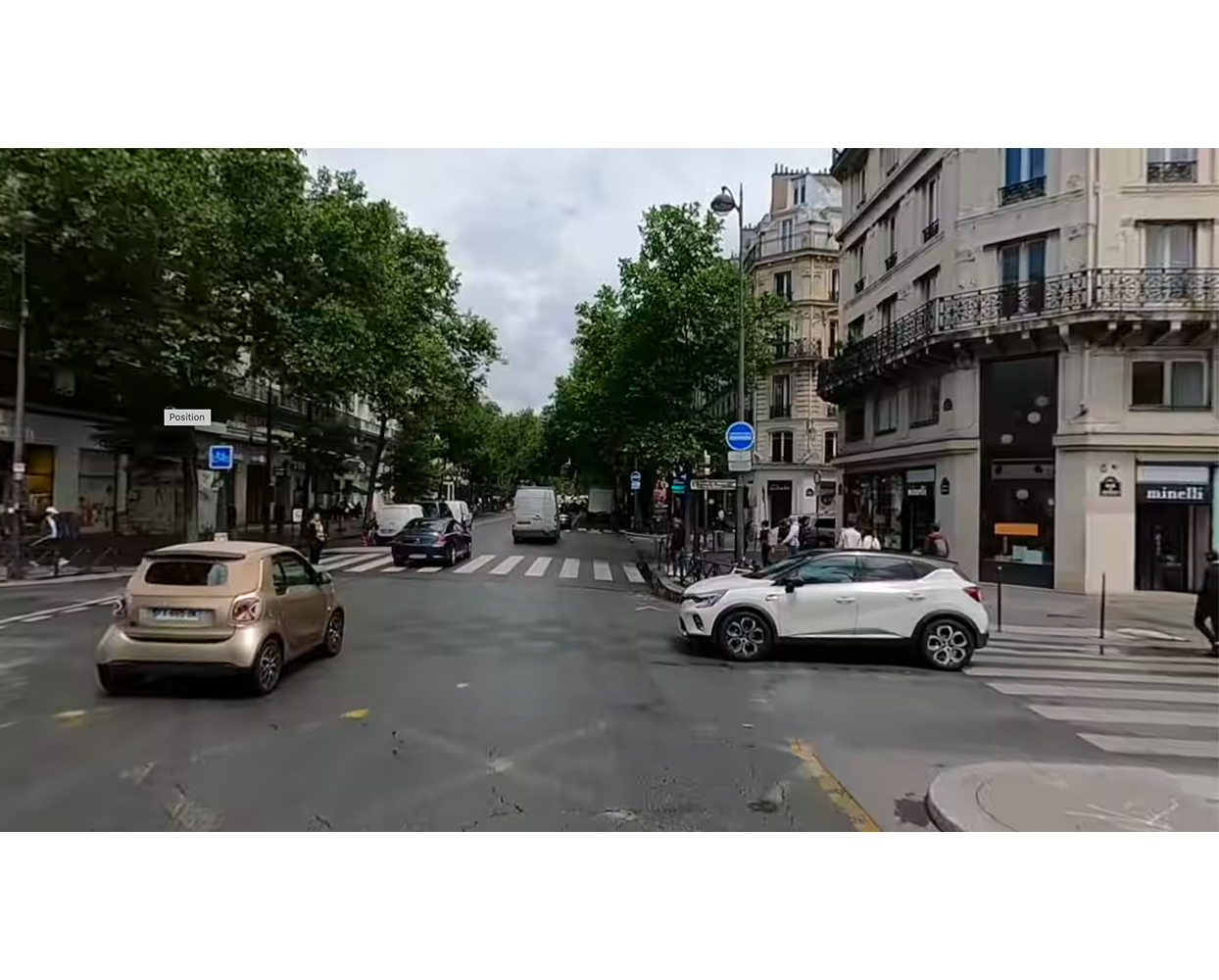}\Description{A few cars are on a street in Paris.}}
 \qquad
 \subfloat[The same frame of video using a modified Little Planet projection~\cite{nguyen_vremiere_2017}.]{\label{fig:360-full-frame}\includegraphics[width=0.4\textwidth]{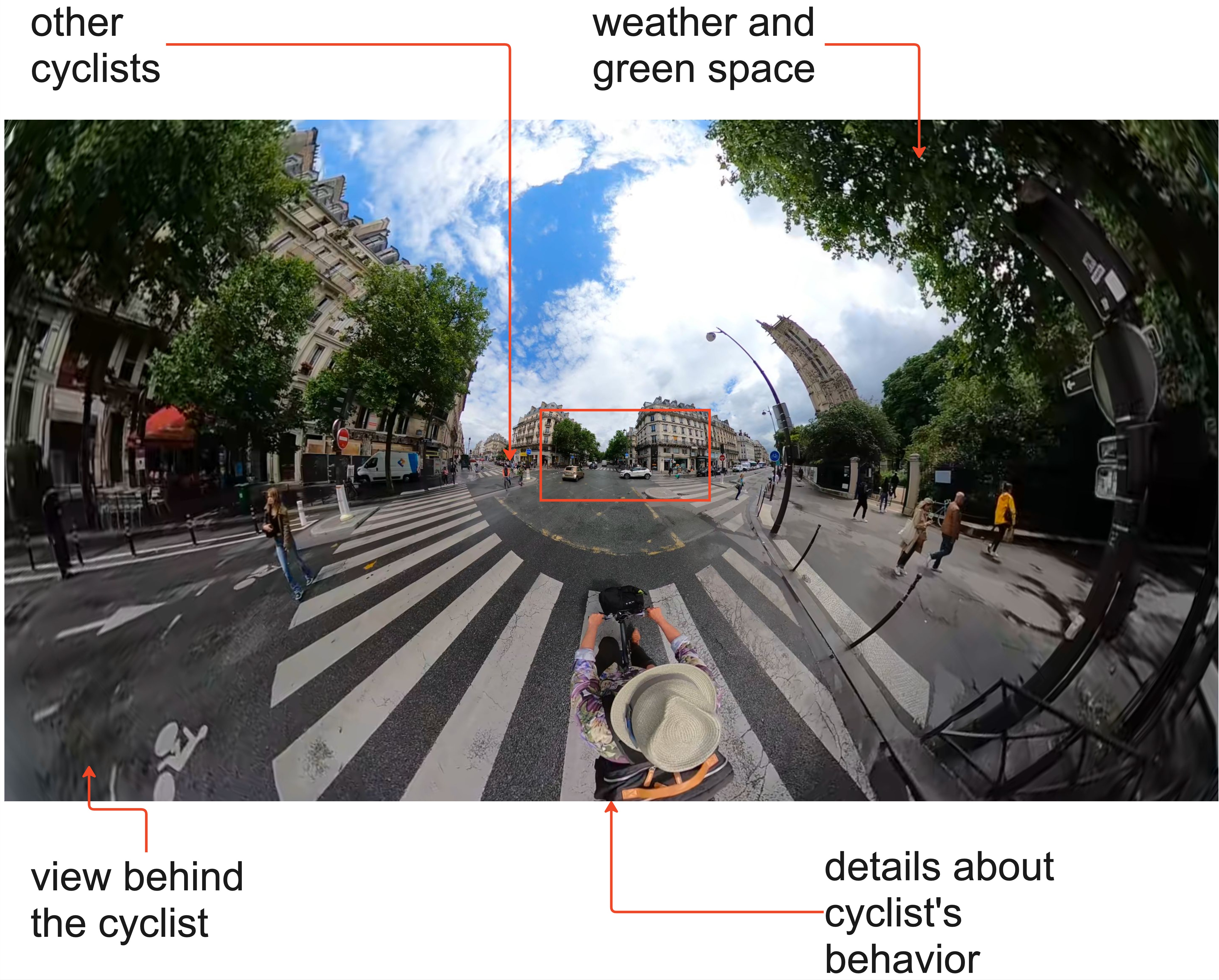}\Description{A distorted picture of someone biking around the street, much more visual information is shown.}}
 \caption{The difference in visual information of a frame when using conventional video and 360° video for an exemplar use case of studying cycling behavior, similar to~\citet{porcheron_cyclists_2023}. Illustrating that 360° images, and videos, (\ref{fig:360-full-frame}) contain significantly more contextual information, at the disadvantage of being significantly more distorted than conventional video. Video from \protect\href{http://www.velomondial.net/}{Velo Mondial} - \protect\href{https://creativecommons.org/licenses/by/4.0/}{CC-BY}.}
 \label{fig:360-why}
\end{figure*}
The process of designers engaging with video as user research material is referred to as Video Design Ethnography (VDE)~\cite{ylirisku_designing_2007}, this iterative process centers around designers viewing and annotating videos individually and then engaging in sense-making to align their understanding of user needs and in turn design goals.
A crucial component that supports this collaborative sense-making are ``video artifacts''\footnote{Not to be confused with compression artifacts, this term has been used by}~{\citet{ylirisku_designing_2007} to describe similar boundary objects made from video.} -- tangible representations of designers' insights, such as storyboards or clusters of screenshots (Section~\ref{bg-conv-video-artifacts}). Designers use the artifacts to represent insights during discussions, documentation, VDE~\cite{buur_ethnographic_2010,ylirisku_designing_2007, nova_beyond_2014, markopoulos_designing_2016} and as the output of the process~\cite{nova_beyond_2014}.

The increasing ubiquity of 360° cameras has the potential to provide designers with richer and more immersive insights~\cite{kramer_innovating_2022}.  With a Field of View (FoV) of 360°, these cameras capture their entire visual context, solving issues with framing~\cite{jokela_how_2019,tojo_how_2021} and enabling viewers to understand more complex interactions -- such as how a cyclist reacts to events in front or behind them~\cite{porcheron_cyclists_2023} or the interaction between the conductor and their orchestra~\cite{vatanen_experiences_2022}. Designers are able to use this additional visual context to gather richer insights~\cite{meijer_sphere_2024} into the context of their users (See Figure~\ref{fig:360-why}).

Unfortunately, as seen in Figures~\ref{fig:teaser-normal} and~\ref{fig:360-full-frame} the spherical nature of 360° video makes it challenging to view and share~\cite{jokela_how_2019} using tools designed around conventional video, such as monitors and video artifacts (Section~\ref{bg-360-video-impact}). To work with 360° video designers need to either discard most of the visual information -- converting it back to conventional video (Figure~\ref{fig:conventional-video}) or suffer from a heavily distorted image (Figure~\ref{fig:360-full-frame}). Thus, in order to take advantage of the benefits of 360° video, it is necessary to create tangible artifacts that enable the kinds of interactions offered by conventional video artifacts~\cite{meijer_sphere_2024}.

In this paper we discuss Tangi, a web-based tool for creating tangible artifacts from 360° video frames, in order to support 360° Video Design Ethnography (Section~\ref{tool}).
To understand the utility of Tangi and the artifacts it creates, we conducted reflection sessions with experienced designers (Section~\ref{generative-sessions}).
These sessions demonstrated that the artifacts Tangi produces enable tangible interactions that~\citet{buur_taking_2000} describe as essential to collaborative video analysis.
Additionally, participants were able to easily modify and create new artifacts using the base elements provided by Tangi, showing the flexibility of paper-based artifacts to evolve to meet the needs of diverse design tasks~\cite{dalsgaard_emergent_2014}.
Finally, we discuss the implications of Tangible 360° Video Artifacts, limitations of this early work, and future steps to further understand how 360° artifacts evolve over a longer design process (Section~\ref{discussion}).

To summarize, this paper's key contributions are:
\begin{enumerate}
  \item Tangi - an open source tool to quickly create Tangible 360° Video Artifacts.
  \item Demonstrating the utility of these artifacts to support collaborative sense-making.
  \item Examples of more complex artifacts that show the ability of paper-based 360° video artifacts to adapt to the needs of specific design teams.
\end{enumerate}

\section{Background and Related Work}
In order to understand the need to develop Tangible 360° Video Artifacts, we will discuss the process of Video Design Ethnography and the function of conventional video artifacts. Then we will discuss the benefits of 360° video for ethnography and, most importantly, how this impacts the creation of video artifacts.
\begin{figure*}[!ht]
    \centering
    \subfloat[The base element of video artifacts, the frame, in this case a frame from Charade~\cite{donen_charade_1963} used by~\cite{goldman_schematic_2006}.]{\label{fig:cva-frame}\includegraphics[width=0.25\textwidth]{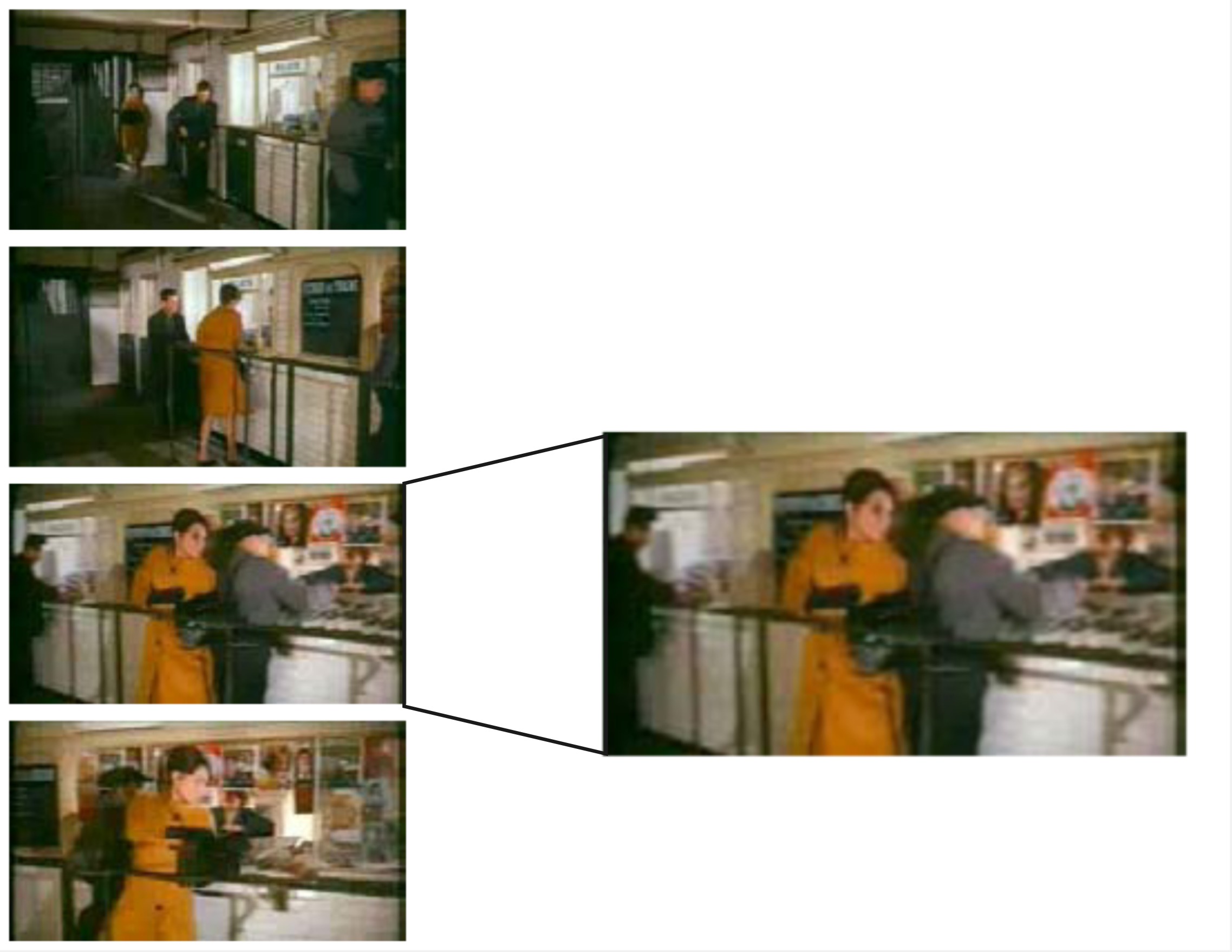}\Description{a sequence of frames from the movie Charade, showing the selection of one frame.}}
    \qquad
    \subfloat[Meta-frame elements; arrows, sticky notes, and speech bubble -- that provide information that does not exist in the frame itself.]{\label{fig:cva-meta-frame}\includegraphics[width=0.25\textwidth]{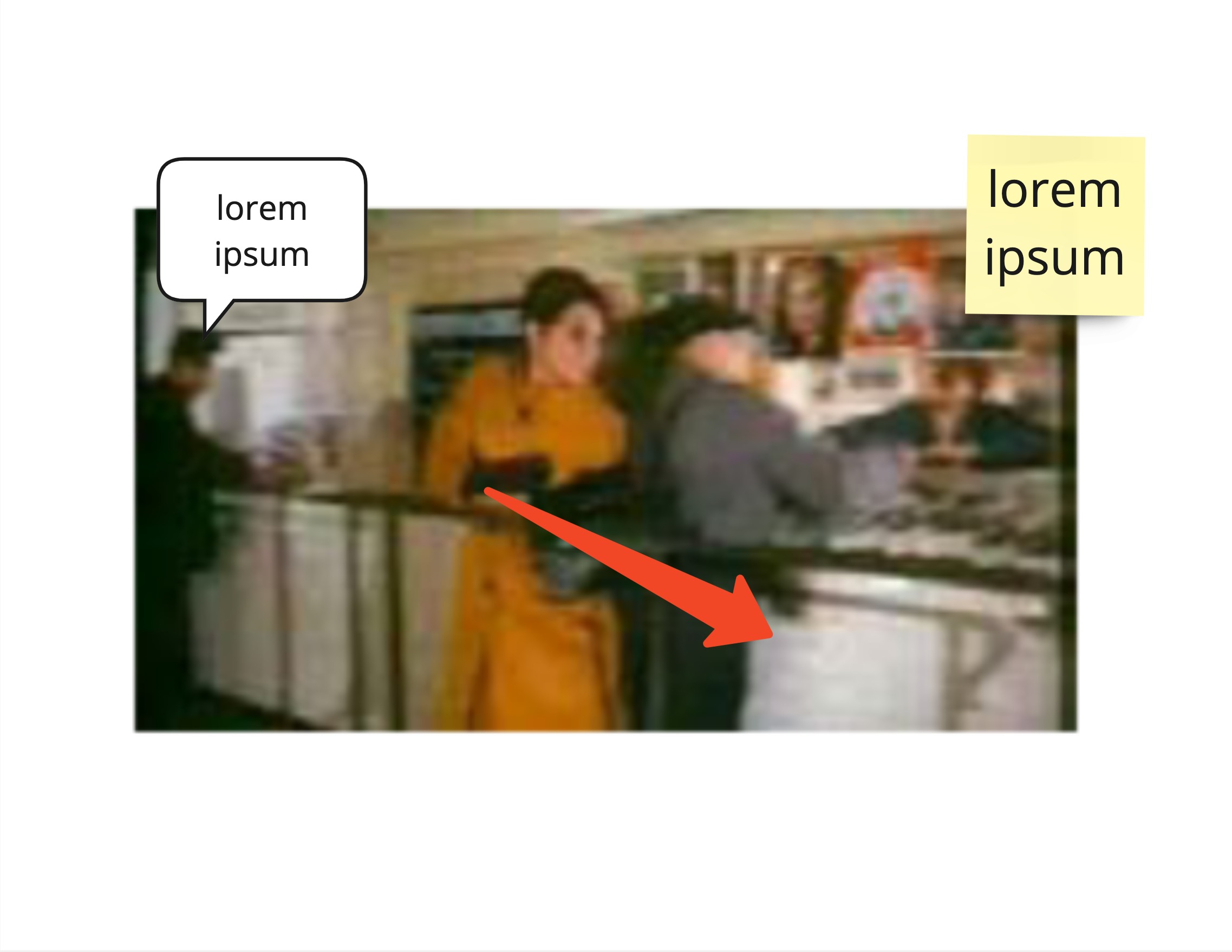}\Description{the frame from figure 4 a with a red arrow, speech bubble, and  sticky note to indicate additional information about the scene.}}
    \qquad
    \subfloat[Multi-frame artifacts combine multiple frames, in this case a cluster of images that feature the theme of ``sneaking'' in various movies\protect\footnotemark.]{\label{fig:cva-multi-frame}\includegraphics[width=0.25\textwidth]{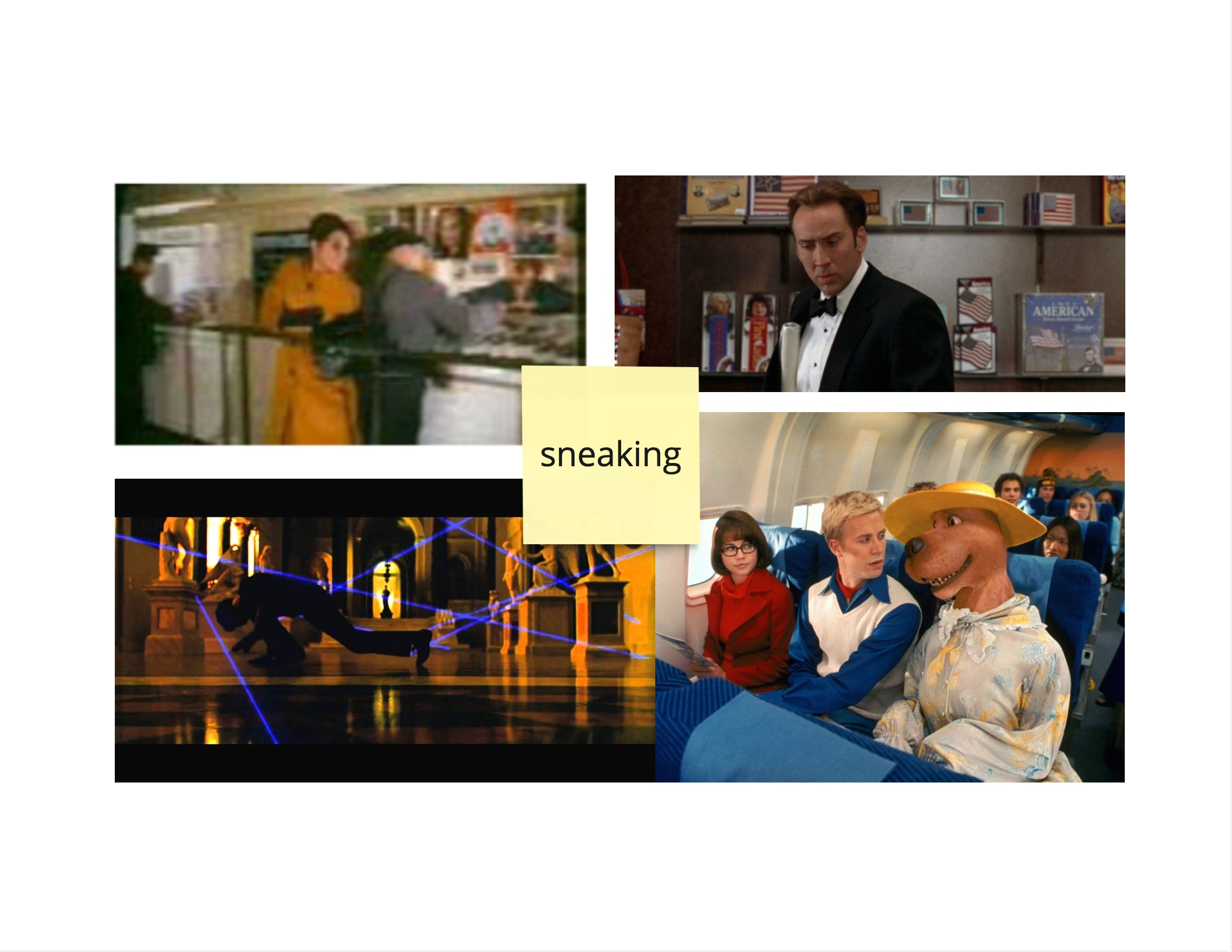}\Description{frames from diverse movies showing the concept of sneaking.}}
    \qquad
    \caption{Examples of the elements of Video Artifacts that designers use or combine to document and share their insights during Video Design Ethnography.}
    \label{fig:conventional-video-artifact-elements}
\end{figure*}
\subsection{Video Design Ethnography}\label{bg-vde}
Design ethnography is the process of how designers and user researchers study (potential) users to gain insights that are used to develop new products and services~\cite{salvador_design_1999}. 
While many forms of information can be used in design ethnography,~\citet{ylirisku_designing_2007} notes that ``video is the medium that conveys most of the detailed richness of a real setting, as compared with text, photos and audio recordings''. 

This rich nature has led to many techniques which use video such as in-situ recordings, video diaries, and interviews~\cite{nova_beyond_2014,ylirisku_designing_2007}, as well as a multitude of methods to analyze video~\cite{vannini_routledge_2020}.

In this work we focus on the process of Video Design Ethnography~\cite{ylirisku_designing_2007,nova_beyond_2014}, which has three general stages~\cite{meijer_sphere_2024}:
\begin{enumerate*}
    \item the initial work of gathering videos,
    \item an iterative analysis process composed of viewing, annotating, and collaborative sense-making, and
    \item finally, sharing the outcome of the process (e.g., design requirements, new insights into context, better understanding of user needs).
\end{enumerate*}
Specifically, the analysis and outcome stages rely on designers overcoming ``...the challenge of conveying their understanding''~\cite[p. 126]{ylirisku_designing_2007} to share their insights.
This is often done using ``video artifacts'' -- boundary objects~\cite{star_institutional_1989} that embody these insights, made up of the video itself.


\subsection{What are Video Artifacts?}\label{bg-conv-video-artifacts}
Video artifacts\footnote{Not to be confused with compression artifacts, this term has been used by}~{\citet{ylirisku_designing_2007} to describe similar boundary objects made from video.} are tangible boundary objects~\cite{star_institutional_1989} created from videos that
--as with most boundary objects-- have many forms which evolve over their use in a design process~{\cite{dalsgaard_emergent_2014}}.
~\citet{goldman_schematic_2006} generalize three types of video artifact -- illustrated in Figure~\ref{fig:conventional-video-artifact-elements} -- the individual frame of a video, meta-frame elements such as arrows and notes, and the arrangement of multiple frames into a single artifact.
These elements can be further combined in complex ways, for example, the multi-frame artifacts can be arranged chronologically (e.g., a timeline~\cite{goldman_schematic_2006}) or categorically (e.g., a mood-board~\cite{nova_beyond_2014}, show in Figure~\ref{fig:cva-multi-frame}) or even form more complex artifacts where timelines are elements on a mood-board.
These categories illustrate the importance of the ``frame'' -- a single moment of a video, embodied as a photo -- as a base element for most video artifacts.
Shifting to 360° video, or 360° photos for that matter, complicates this base element of the frame, creating one of the major barriers for adopting 360° video in design ethnography (see Section~\ref{bg-360-video-impact}).

\footnotetext{Movies shown from top left, clockwise: Charade (1963), National Treasure (2004),  Oceans 12 (2004), and Scoobie-Doo The Movie (2002).}
\subsection{Importance of Tangibility}\label{bg-tangibility}
While many interactions with video are digital (editing, viewing, etc.), \citet{buur_video_2000} specifically point to the critical use of tangible video artifacts in collaborative sense-making to prevent the interruption of digital tools``[...] into the social sphere of design discussions without restraining the dynamics''.
Additionally,~\citet{brandt_how_2007} describe how tangible artifacts provide designers with ``things to think with'', framing and aligning discussions and analysis by presenting themselves as a tangible token of abstract ideas.~\citet{markopoulos_designing_2016} point to the importance of using tangible cards in the video card game method defined by~\citet{buur_video_2000}, because these cards:
\begin{enumerate*}
    \item afford important manipulations such as pointing, rotating, or arranging,
    \item can be marked on and annotated to record the discussion and create meta-frame elements in real time,
    \item and trigger combinatorial creativity.
\end{enumerate*}
It is precisely due to the tangible nature of these artifacts that theysupport the highly collaborative sense-making processes that designers engage in.
When being used to communicate the results of such a process (i.e., the output stage of VDE), the tangibility of video artifacts has an additional benefit: it makes them persistent~\cite{markopoulos_designing_2016}. This persistence is an important factor of the video artifact acting as a boundary object and providing different stakeholders with a common frame of reference~\cite{brandt_how_2007} and avoiding the ambiguity of a video clip~\cite{markopoulos_designing_2016} by selecting a single, persistent frame. Finally, this persistence is indicative of the ``artifact'' nature of a video artifact, making it easier to preserve and allowing designers to re-engage with the knowledge embodied by video artifacts from previous projects or design teams~\cite{ylirisku_designing_2007}.

\subsection{Why Use 360° Video for Design Ethnography}
The distinguishing factor of 360° cameras is that they capture the entire visual context around the camera at once. This has numerous advantages -- on a base level it removes the challenge of ``framing'' the scene in the view of the camera~\cite{jokela_how_2019,tojo_how_2021}. This is particularly beneficial for VDE, where the focus of the analysis (and thus where the camera is pointed) changes throughout the exploratory process~\cite{ylirisku_designing_2007}. Figure~\ref{fig:360-why} illustrates how 360° video allows the viewer to explore different interactions and actors within a space that is not captured by conventional video. This also means that 360° video provides viewers with the ability to observe specific parts of complex, multi-actor interactions -- such as students and a teacher~\cite{kosko_using_2022}, a conductor and an orchestra a~\cite{vatanen_experiences_2022}, or how a cyclist reacts to their environment~\cite{porcheron_cyclists_2023,meijer_sphere_2024}.
Additionally, the immersive qualities of 360° video enables designers to engage in immersive qualitative analysis~\cite{vatanen_experiences_2022}, grounding insights in context, and leading to greater empathy~\cite{pimentel_voices_2021,shin_empathy_2018}. Thus, 360° video makes VDE easier (by removing the challenge of framing the video) and the insights richer (by enabling viewers to connect multiple actors and engage in more empathic analysis of the video) simply because it captures the full visual context around the camera.

\subsection{How 360° Video Complicates Video Artifacts}\label{bg-360-video-impact}
The core challenge of creating tangible 360° video artifacts is the complex nature of a frame of 360° video.
When presented as a flat artifact\footnote{Necessitated by many of the ways we document and share information, such as this paper.}, a full frame of 360° video, or a 360° photo for that matter, is distorted (Figure~\ref{fig:360-full-frame}, making understanding the visual information in the scene -- especially spatial relations within that frame~\cite{koenderink_planispheric_2017} -- cognitively challenging~\cite{xiao_recognizing_2012,jokela_how_2019}.
To avoid this issue, one can use a perspective frame -- a subset of the full 360° frame with a conventional FoV -- resulting in a conventional screenshot without distortion but entirely removing the ``360°'' nature of a 360° video. Conversely, designers could create artifacts with larger than normal FoVs but not the entire 360° frame -- resulting in trade-offs between distortion and context.
This is further complicated by the fact that 360° video has the potential to generate insights based on connections between disparate areas of a 360° frame~\cite{meijer_sphere_2024} (visualized in Figure~\ref{fig:360-full-frame}), adding another challenge to creating a 360° video artifact.
While these distortions could be addressed using technology such as VR headsets or spherical displays, both of these have the potential to reintroduce the challenges that make \textit{tangible} artifacts so important (see~{\ref{bg-tangibility}}).
VR headsets can isolate the viewer~{\cite{neubauer_experiencing_2017}}, again breaking the discussion discussed by~{\citet{buur_video_2000}}.
For example, while spherical displays enable an undistorted view of 360° video that could enable collaborative interaction they require complex setups with external projectors~{\cite{li_omnieyeball_2016}}, which limits when and where they can be used. Crucially, these digital interventions lack the persistence, arrangability, and ease of modification offered by paper artifacts that are important to VDE~{\cite{markopoulos_designing_2016}}.

\section{Design Space of 360° Video Artifacts}\label{design-considerations}
\begin{figure}[!t]
    \centering
    \includegraphics[width=0.6\linewidth]{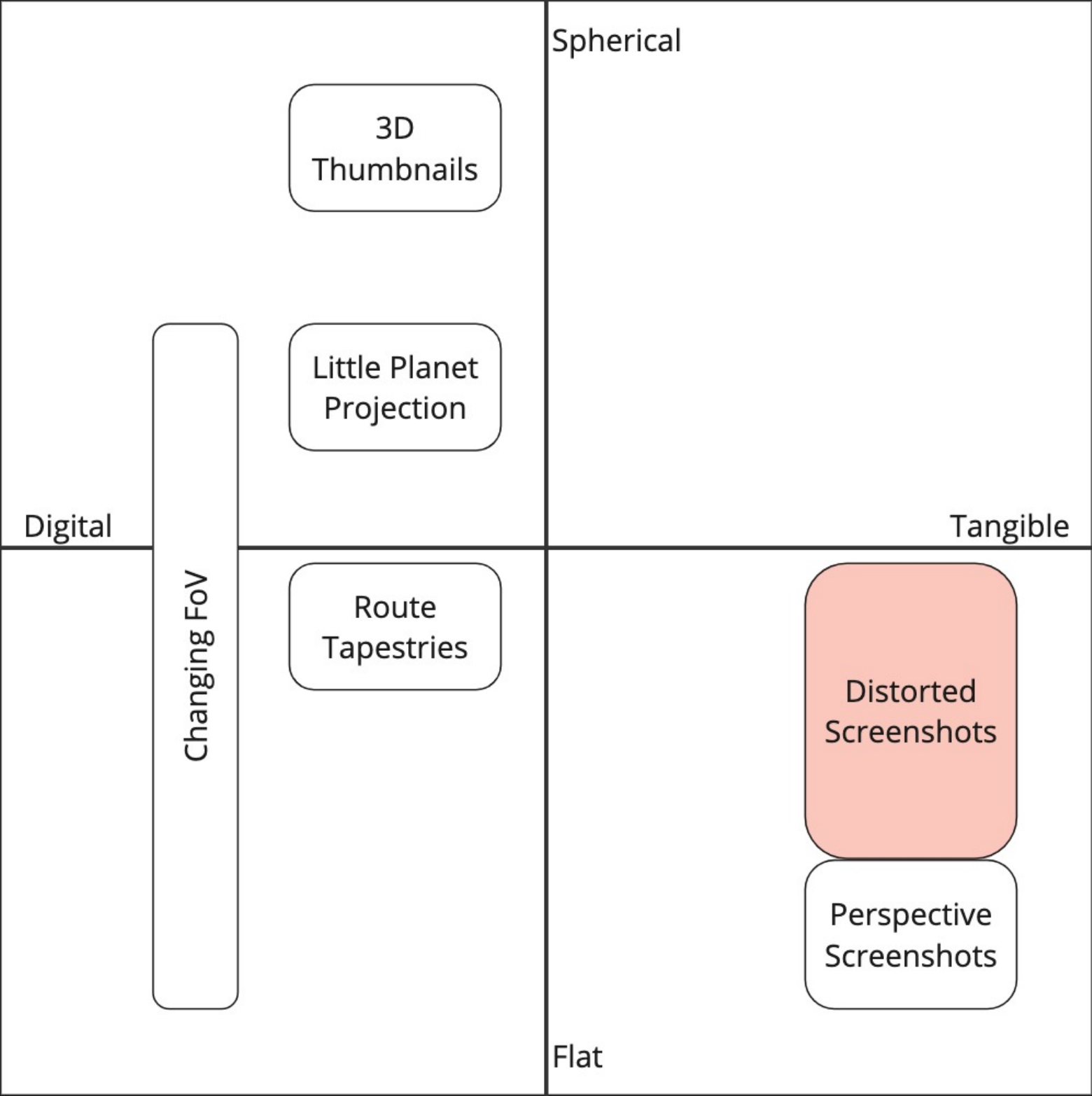}
    \caption{The design space of 360° video artifacts with examples of different approaches of interacting with 360° content indicating the gap for Tangi: the lack of tangible artifacts that contain the full 360° context.}
    \Description{a design space as a 2D graph with the axes of spherical-flat and digital-tangible. On the left (digital) side there are examples of digital artifacts for 360° content, on the right there are two boxes in the lower (flat) corner showing perspective screenshots and distorted screenshots.}
    \label{fig:design-space}
\end{figure}
To further motivate and illustrate the challenge faced by designers wishing to engage with 360° VDE, we sketch out the design space (Figure~\ref{fig:design-space}) for 360° video frame artifacts, defined by two axes:
\begin{enumerate}
    \item \textbf{Spherical -- Flat:} how much the artifact matches the spherical nature of 360° video, where the artifacts at the bottom only encapsulate a cropped subsection of the full sphere of 360° video, and the top end a complete, non-distorted, sphere. 
        \end{enumerate}
\begin{enumerate}[resume]
    \item \textbf{Digital -- Tangible:} the embodiment of the artifacts, where fully digital artifacts leverage interactivity (e.g., changing FoV or perspective of the video) to reduce distortion and/or show additional visual context -- and fully tangible artifacts support the interactions described in Section~\ref{bg-tangibility} and are thus physical and persistent.
\end{enumerate}

\begin{figure}[!h]
    \centering
    \includegraphics[width=0.99\linewidth]{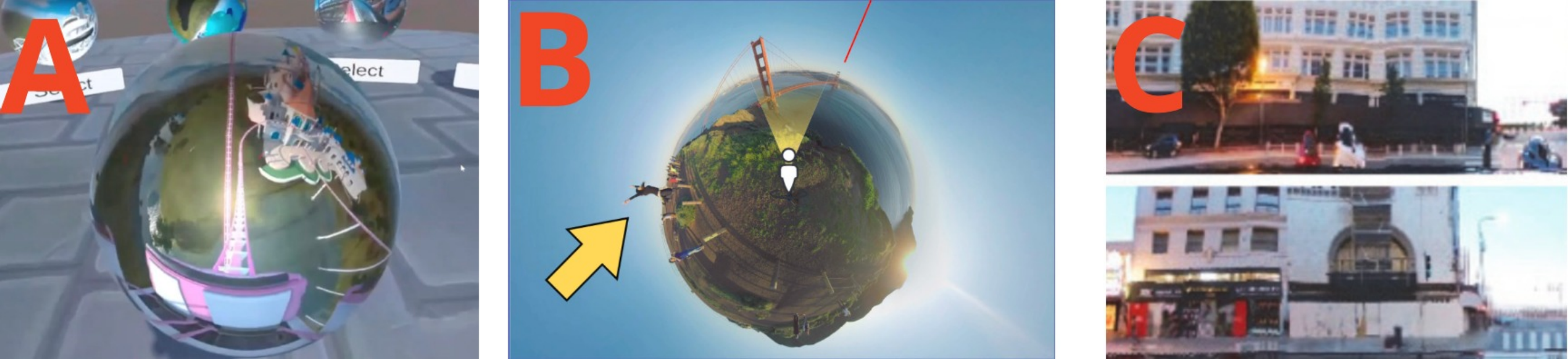}
    \caption{Examples of (A) 3D Thumbnails~\cite{vermast_introducing_2023}, (B) Little Planet~\cite{nguyen_vremiere_2017}, and (C) Route Tapestries~\cite{li_route_2021}.}
    \label{fig:different-tools}
    \Description{3 examples of 360° artifacts, the first is a 3D thumbnail -- a sphere that someone in VR could move around. Second the little planet projection which shows a distorted overview of the full scene. Thirdly route tapestries which show the two side views of someone traveling down a street.}
\end{figure}
Figure~\ref{fig:design-space} also shows how previous approaches of sharing 360° content fit into the design space. The first example is 3D Thumbnails~\cite{vermast_introducing_2023} (Figure~\ref{fig:different-tools}-A) which enable users in VR to get an overview of 360° content by creating a spherical screenshot that the viewer can move around -- thereby including the full context, but only when the interaction is digital. The second is the Little Planet projection  (Figure~\ref{fig:different-tools}-B) discussed by~\citet{nguyen_vremiere_2017}, which provides a view with a lot of the visual context but with high distortion -- counteracted by a second view in a VR headset. Finally,~\citet{li_route_2021} discuss Route Tapestries (Figure~\ref{fig:different-tools}-C) which provide a view of the sides of a 360° video to provide an overview of 360° videos -- again this is tied to a 360° video player with a conventional FoV. These approaches all rely on a digital interaction (controlling the perspective of a video player with a conventional FoV), making them not ideal for design workshops.

On the physical side of the design space, there are screenshots of 360° content -- either by cropping the 360° video or using a heavily distorted projection of the 360° video onto a flat surface.
\begin{figure*}[!ht]
    \centering
    \subfloat[Flat 360° video artifact.]{\label{fig:flat-artifact}\includegraphics[width=0.35\textwidth]{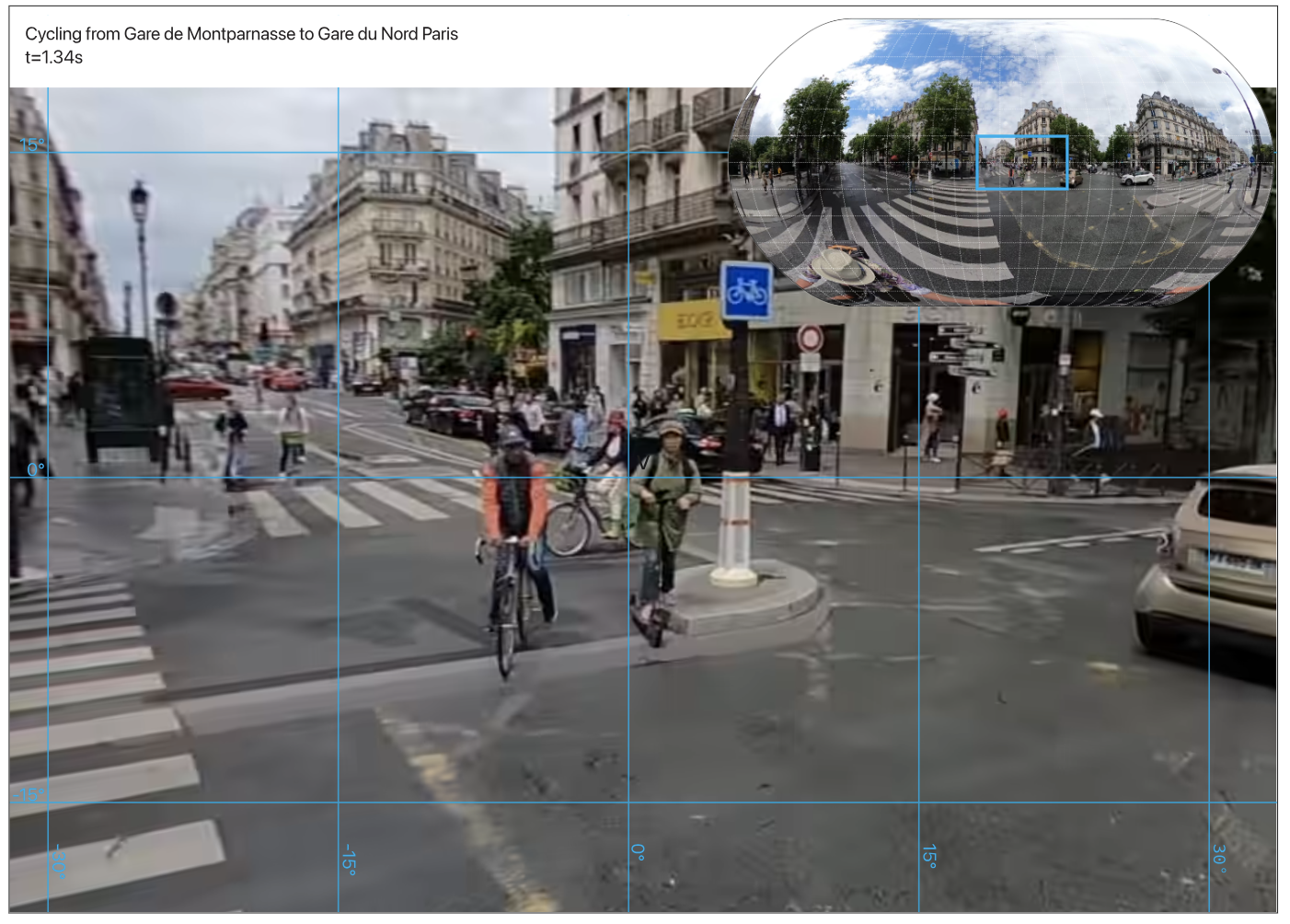}\Description{ a piece of paper with a perspective crop of a larger 360° image, with the full 360° image as a map projection shown in the upper right hand corner.}}
    \qquad
    \subfloat[A ``cube'' Sphere-ish artifact.]{\label{fig:cube-artifact}\includegraphics[width=0.25\textwidth]{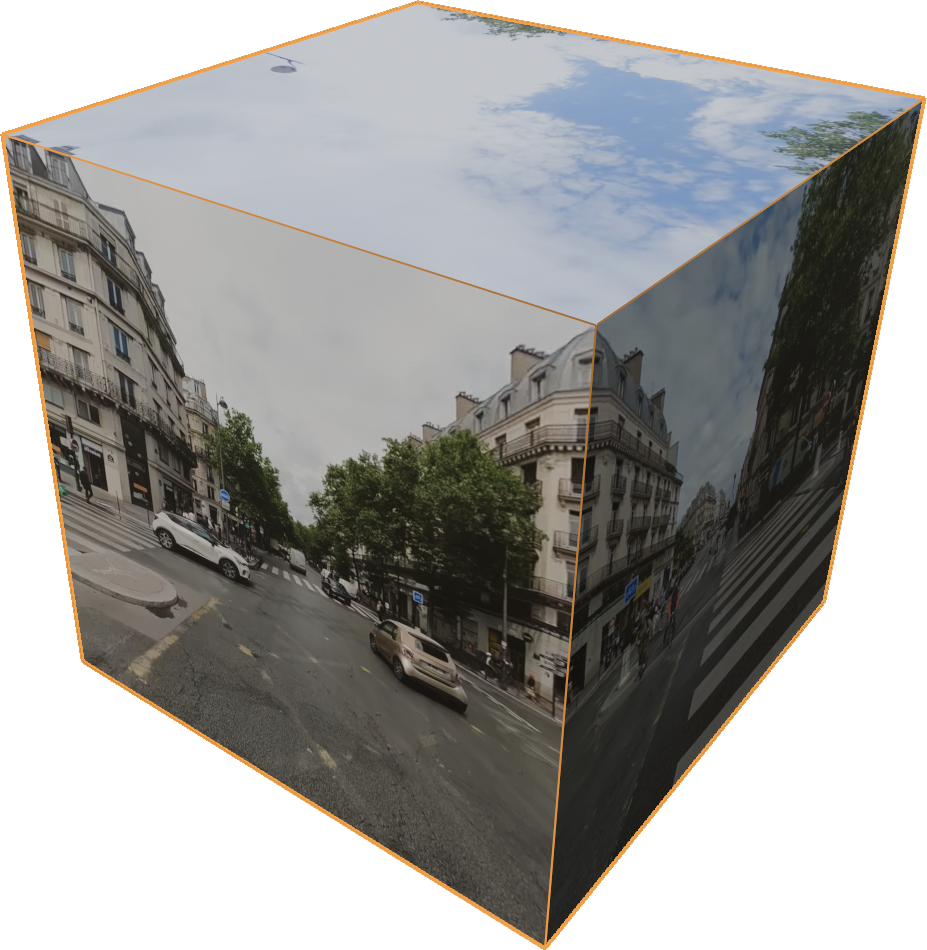}\Description{the same overall 360° image projected onto a cube.}}
    \caption{Examples of the two types of the Tangible 360° Frame artifacts generated by Tangi. Video from \protect\href{http://www.velomondial.net/}{Velo Mondial} - \protect\href{https://creativecommons.org/licenses/by/4.0/}{CC-BY}.}
    \label{fig:initial-artifact}
\end{figure*}
\subsection{Design Considerations}
Given the importance of tangibility and the purpose of conventional artifacts(Section~\ref{bg-conv-video-artifacts}), we define three design considerations for 360° video artifacts for VDE:
\begin{enumerate}
    \item \textbf{Tangibility:} As discussed in Section~\ref{bg-tangibility}, the tangibility of video artifacts supports collaborative engagement with insights from the video. This tangibility also supports provides the practical benefits of creating easy-to-modify artifacts~\cite{markopoulos_designing_2016} that are essential in enabling the collaborative negotiation process of VDE{~\cite{buur_ethnographic_2010}}. Therefore, 360° video artifacts for VDE should be tangible.
    \end{enumerate}
\begin{enumerate}[resume]
    \item \textbf{Retain 360° Context:} One approach to address the challenge of 360° video artifacts is to simply crop the video, turning it into a conventional video that can be analyzed and documented using conventional video artifacts. However, this also throws away most of the visual context -- which is the benefit of 360° video. Therefore, a 360° video artifact should retain the full visual context of 360° videos.
        \end{enumerate}
\begin{enumerate}[resume]
    \item \textbf{Minimize Distortion:} fundamental to the challenge of just using full 360° frames as the basis of an artifact is the complexity of understanding the distorted image~\cite{xiao_recognizing_2012,jokela_how_2019,koenderink_planispheric_2017}. Therefore, to support their use when sharing insights, 360° video artifacts should minimize distortion of the 360° image.
\end{enumerate}

\section{Tangi}\label{tool}
Given the need specific design requirements discussed in Section{~\ref{design-considerations}}, we developed Tangi, a tool that enables designers to rapidly create tangible artifacts from 360° video. Based on the design space shown in Figure{~\ref{fig:design-space}}, we developed two distinct approaches to making frame level artifacts, both of which can can be quickly created with Tangi's online interface.

\subsection{Two Approaches to Tangible Frames}
We started the exploration of possible artifacts by focusing on paper-based approaches~\cite{he_data_2024} to create artifacts tangibility and ease of modification pointed to in Section~\ref{design-considerations}. We then turned to works of cartography\footnote{a field that often deals with presenting spherical information using paper-based artifacts.}~\cite{ptolemaeus_geographia_1845,maceachren_how_2004,lee_nomenclature_1944} which pointed to two approaches that bridge the design space set up in Section~{\ref{design-considerations}}. 
In the design space shown in Figure~\ref{fig:design-space}, \textbf{flat artifacts} enhance perspective screenshots by adding contextual information (i.e., moving up), and sphere-ish artifacts move spherical visuals to the tangible domain (i.e., moving right).
\\
\textbf{Flat 360° Video Artifacts}
enable designers to create 360° video artifacts on the same flat paper as conventional video artifacts, minimizing the complexity of adopting 360° video in VDE. As shown in Figure~\ref{fig:flat-artifact}, this was done by using a combination of non-distorted perspective screenshot and a mini-map that includes the full visual context and provides orientation of the perspective. This approach of providing a more distorted overview was inspired by atlases, which also provide the concept of graticules~\footnote{A grid of dotted lines to provide coordinate information in maps. See: \url{https://en.wikipedia.org/wiki/Graticule_(cartography)}.} to help orient the screenshot in 360° space.

\textbf{Sphere-ish 360° Video Artifacts}
provide the entire 360° frame in an approximation of the actual spherical nature of the video. While this class is inspired by globes, making perfectly spherical globes is a time-consuming process~\footnote{See: CBS How Are Globes Made : The Art of Making Globes \url{https://youtu.be/d0Lyw42Klew?t=48}.}. Thus, we experimented with a variety of polyhedra to quickly make ``sphere-ish'' 3D representations of 360° video frames preliminary evaluation within the research team resulted in two shapes being selected:
    \begin{enumerate}
        \item the cube (Figure~\ref{fig:cube-artifact}, printable example~\ref{fig:example-cube}). Simple to cut and fold, and also simple to understand (given the familiarity of a box).
        \item the cuboctahedron~\footnote{See:~\url{https://en.m.wikipedia.org/wiki/Cuboctahedron}. It also shares similarities with the work of~\citet{hurbain_3-d_2003}, a wonderful example of tangible 360° photo artifacts from at least 2003 -- unfortunately not explored as a tool for design ethnography.} (shown in Figures~\ref{fig:teaser-poly} and~\ref{fig:tool-sphere-ish}, printable example~\ref{fig:example-poly}). It has more facets than the cube, but importantly retains 6 large, orthagonal faces.
    \end{enumerate}
\subsection{Tangi: the Online Tool}
\begin{figure}
    \centering
    \includegraphics[width=1\linewidth]{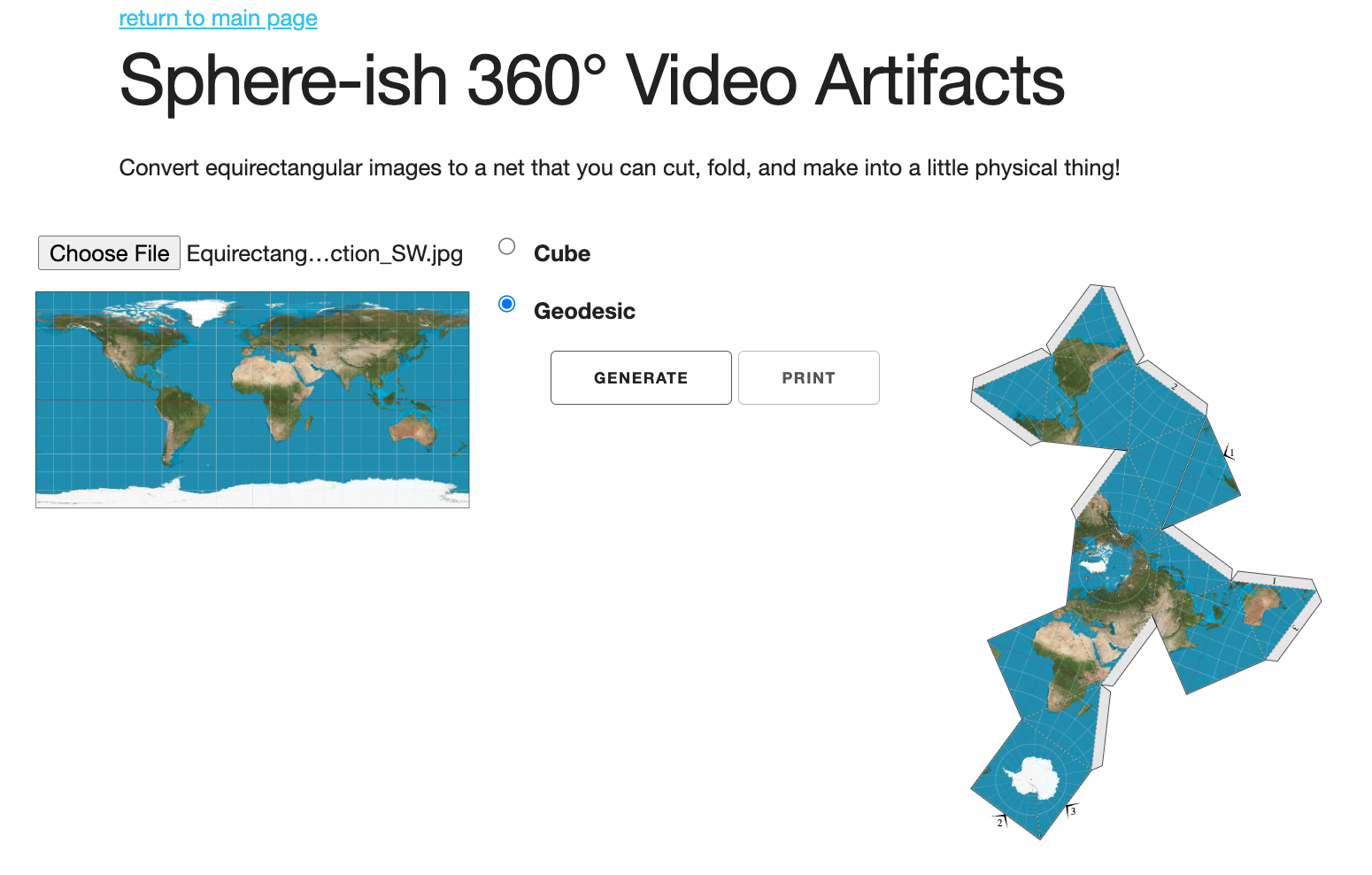}
    \caption{The generator for sphere-ish artifacts, showing an example of generating a globe using an equirectangular map.}
    \label{fig:tool-sphere-ish}
    \Description{a screenshot of a web-page that shows a flat projection of the earth on the left and the same image projected on the flat pattern of a cuboctahedron on the right.}
\end{figure}

The online component of Tangi converts equirectangular 360° screenshots\footnote{which can be taken by 360° video players such as VLC~\url{https://www.videolan.org/}} into either flat or sphere-ish artifacts.
The landing page
briefly describes the two types of artifact and provides links to generate them. The sphere-ish artifact generator (Figure~\ref{fig:tool-sphere-ish}) allows the user to simply open a 360° image file, select between a cube or cuboctahedron. It then projects the image onto the flat cut-and-fold template which can then be printed. Similarly, the flat artifact generator allows the user to upload an image, select an area of interest, and generate a flat artifact as shown in Figure~\ref{fig:flat-artifact}. The tool is available at [URL removed for review].

While Tangi is not the first tool that creates paper models of 360° images~\cite{hurbain_3-d_2003,noauthor_saya360_2017,yoshida_how_nodate}, the focus of Tangi is to specifically support designers by creating both flat and sphere-ish artifacts.
Regardless of the tool used to generate them, this paper focuses on the utility of tangible 360° video artifacts for design work, and we encourage the community to expand on Tangi by modifying the source code [URL removed for review] or creating their own tools.
\section{Methodology}\label{generative-sessions}
In order to understand the utility of Tangi and the artifacts it generates, we conducted sessions with nine designers who had experience with both 360° video and video design ethnography. We used snowball sampling to recruit participants (shown in Table~\ref{tab:participants}) that to conduct a 2-3 hour in-person sessions with the lead researcher. The sessions addressed the following questions:
\begin{enumerate}
    \item How do the example artifacts enable the tangible interactions designers use during VDE?
    \item What are functions of tangible artifacts that are unique to 360° video?
    \item How can the artifacts generated by Tangi support the creation of more complex and bespoke artifacts?
\end{enumerate}

\begin{table}[H]
\begin{tabular}{lll}
Participant &  Design Experience & 360° Video \\ \hline
P1 & 8 years & 1 year \\
P2 & 15 years & 10 years \\
P3 & 15 years & 5 years \\
P4 & 6 years & 2 years \\
P5 & 10 years & 3 years \\
P6 & 8 years & 1 year \\
P7 & 2 years & 2 years \\
P8 & 3 years & 2 years \\
P9 & 6 years & 1 years 
\end{tabular}
\caption{The relevant experience of the participants who engaged in the expert evaluations.}
\Description{Table 1: a simple table that lists 9 participants (P1-P9), their experience with design in years and experience with 360° video in years.}
\label{tab:participants}
\end{table}

\subsection{Session Setup}
To simplify the challenges of coordinating working designers, the 2 to 3 hour sessions were conducted one-on-one with a participant and the lead researcher. The lead researcher acted as a fellow designer who shared a number of insights from a 360° video using artifacts generated by Tangi (example artifacts included in Appendix~\ref{apx:example-artifact}). In addition to these initial artifacts, the lead researcher had a collection of 360° videos (listed in Appendix~{\ref{apx:example-videos}}) which participants could use to create additional artifacts. Participants were provided with workshop materials (sticky notes, pens, paper, dots, etc.) and paper-craft tools (cutting mat, box cutter, scissors, rotary perforation, etc.).

\subsection{Session Flow}
\begin{enumerate}
    \item The lead researcher explained the purpose and elements of tangible video artifacts for conventional video using Figure~\ref{fig:conventional-video-artifact-elements}, as well as the text from Section~\ref{bg-conv-video-artifacts} for reference.
    \item Participants were introduced to the specific challenge that 360° video introduces to creating these kinds of video artifacts using the text of Section~\ref{bg-360-video-impact}.
    \item Participants were asked to reflect on their own work and challenges with 360° video in general, and with creating artifacts specifically.
    \item The lead researcher shared a number of example 360° videos (Appendix{~\ref{apx:example-videos}}) and pre-assembled artifacts (Appendix{~\ref{apx:example-artifact}}) to demonstrate the artifacts created by Tangi.
    \item Using these artifacts, participants were guided in a reflection on the concept of utility of the artifacts, contrast how the different classes of artifact provide different functions for collaboration, and when and how they might use these artifacts in their own work with 360° video.
    \item During this reflection, participants were encouraged to create new artifacts and to modify artifacts to explore their ideas of how to adapt them to their needs.
\end{enumerate}

\subsection{Data and Analysis}
Each session was recorded using a voice recorder, documented in a research notebook, and photos were taken of participants' notes, sketches, and artifacts they modified or created. We analyzed this data using reflexive thematic analysis~\cite{braun_using_2006,braun_thematic_2022} with an inductive process.
\begin{enumerate*}
    \item The lead researcher spent two weeks familiarizing themselves with the data, reviewing sketches, notes, recording transcripts, and the artifacts created by participants.
    \item An initial set of codes were generated from the data, focused on understanding the \textit{utility}~\cite{ledo_evaluation_2018} of the artifacts for participants, as well as comparing the capabilities of 360° artifacts to conventional video artifacts~\cite{buur_taking_2000,ylirisku_designing_2007,markopoulos_designing_2016}.
    \item After checking these codes for consistency, clarity, and completeness, the research team iteratively created categories and re-engaged with the data to identify emergent themes -- a process akin to VDE (Section~\ref{bg-vde}).
    \item In a final meeting, the research team discussed the resulting themes, finally selecting and defining the themes defined and described below.
\end{enumerate*} 

\section{Results}\label{results}
Overall, participants echoed the need for tangible 360° video artifacts to support collaborative analysis expressing that the tool was \textit{``...very cool - mhmm- really cool''}~(P4) and that the artifacts created by Tangi would \textit{``let me share my thoughts''}~(P9). Specifically, Tangi's simple interface and paper-based artifacts impressed participants with its \textit{``...super low threshold to produce and create''}~(P6) which enabled the lead researcher and participants to create 34 artifacts over 9 sessions. After the initial sessions, participants 1, 5, 6, and 9 experimented with Tangi after the sessions for personal and professional design work.
The sessions demonstrated that artifacts created by Tangi supported alignment and discussions in a similar way to tangible artifacts for conventional video~{\cite{markopoulos_designing_2016,buur_ethnographic_2010,buur_video_2000}} without the difficulties of using conventional artifacts for 360° content~{\cite{meijer_sphere_2024}}.

As none of the participants mentioned significant issues with the utility of the online component of Tangi, our analysis focuses on the artifacts it generates.
Our analysis revealed four themes specific to the utility of the artifacts generated by Tangi for the kinds of collaborative sense-making processes found in VDE:
\begin{enumerate}
    \item \textbf{D}ifferences between the different classes of artifacts,
    \item \textbf{F}unctions specific to tangible \textit{360°} video artifacts,
    \item \textbf{M}oving beyond the ``frame'' element,
    \item \textbf{T}angibility and time.
\end{enumerate}
Below we describe these themes and include quotes as well as images of the artifacts created during the sessions and visualizations of participants' actions.

\noindent\textbf{[D]ifferences between the different classes of artifacts}
\begin{enumerate}[label= D\arabic*]
    \item \label{result:difference-flat-sphere-ish}\textbf{Difference between flat and sphere-ish artifacts:} one primary distinction was the impression they gave to the observer, from the flat artifacts giving \textit{``a more clinical, analytical perspective compared with the immersive nature of the cube''}~(P5). In more extreme terms, the sphere-ish artifacts made P6 reflect that they \textit{``...were God or something, I could see this complex interaction between all of these things''}. Other labels used to compare the classes of artifacts included \textit{``conscious and unconscious''}~(P1), \textit{``human and more than human''}~(P3), and \textit{``atmospheric and analytical''}~(P2).
    
    This distinction also influenced how and when participants would use the different artifacts. For example P6 had a clear preference for the sphere-ish artifacts to \textit{``provide an overview of the space, so if I am an architect looking at how to redevelop a space I would love that''}, strongly connecting the complete context with an overview of a space. P5 reflected on how these needs change through the design and analysis process and that by transitioning between flat and sphere-ish artifacts could enable a process \textit{``...of inhabiting and then what's the word like observing? ... switching between the first person and analytical perspectives''}.
\end{enumerate}
\begin{enumerate}[resume,label= D\arabic*]
    \item \label{result:difference-cube-octaherdon}\textbf{Difference between cube and cuboctahedron:} Participants noted that the cube allowed one to only see one face at a time, which gave more of a sense of \textit{``...simply 6 flat images, which is more useful for analysis''}~(P5). In contrast, the cuboctahedron was perceived more like a sphere because \textit{``I cannot just look at 1 [face]. I will immediately already see multiple, so I am reminded that this is not just a weird picture in 2D''}~(P9).

    P1, P6, P8, and P9 all described how the smooth transition between the facets of the cuboctahedron also \textit{``...encourages you to keep rotating around the sphere, you don't know when to stop''}~(P1), which was seen as an advantage. However, P8 argued that this \textit{``...makes it more difficult to understand since you have so many faces and you move from one to the next without stopping''}. P3, P4, and P7 similarly preferred the simplicity of the cube over the cuboctahedron, P2 and P5 not expressing a clear preference.
\end{enumerate}

\noindent\textbf{[F]unctions Enabled by Tangible 360° Video Artifacts}
Based on their interactions with the artifacts, participants described -- and created their own artifacts -- that touch on two functions that are unique to tangible video artifacts based on 360° content: the need to orient their understanding of the frame, and creating connections between the details of an insight and the overall 360° visual context.
\begin{enumerate}[label= F\arabic*]
    \item \label{result:orientation}\textbf{Orientation:} Participants noted that one of the important challenges when working with 360° video was ``orientation'', both how \textit{``the spatial relations between people in a single 360° image is impossible to understand''}~(P3) and understanding the orientation of the video over a long time. Here, the facets~\footnote{The large, flat faces.} of the sphere-ish artifacts provided a clear set of sides to help multiple participants in a workshop agree on a specific orientation (P1, P5, P6, P7) and frame discussions during analysis (P1, P4, P6, P8). Specifically, P2 noted that the choice\footnote{In reality this was a fortunate coincidence of the orientation of the video, which was helpful in eliciting this aspect of 360° video artifacts.} to orient the major facets of the sphere-ish artifacts with the road in the example artifact was a smart choice that would simplify how multiple participants would understand the context as well as simplify the alignment of multiple artifacts of different road use.
\end{enumerate}
\begin{enumerate}[resume,label= F\arabic*]
    \item \label{result:context-detail-link} \textbf{Context-Detail Link:}
    One of the main functions discussed by participants was how 360° video creates a need to link the overall visual context to specific details   In away this emerged from the creation of the flat artifacts which include the overall context in the form of the mini-map. Crucially this was missing in the sphere-ish artifacts which provided the entire context, however P5 noted: \textit{``I'm not sure how to point out one specific thing. I feel like I want to, almost, desaturate the whole thing except the important part''}. Many participants opted to address this shortcoming by creating sphere-ish meta-frame elements (\ref{result:sphere-meta-frame}) discussed below.
\end{enumerate}

\noindent\textbf{[M]oving beyond the ``frame'' element}
While the artifacts presented to participants were initially limited to frame elements, which is commonly used for analysis of conventional video~{\cite{markopoulos_designing_2016,buur_ethnographic_2010,buur_video_2000}}, participants' reflections expanded to include examples of meta-frame\footnote{Elements such as text and drawings that do not exist within the body of the frame. See Section~\ref{bg-conv-video-artifacts}.} annotations for sphere-ish artifacts, multi-frame artifacts that combine both types of artifacts, and several ideas that were important to mention.

\begin{figure*}[!ht]
    \centering
    \subfloat[The meta-frame annotation booklet connected to a cube artifact by string, created by P6. The label ``sky'' is used to orient discussions (see~\ref{result:orientation}).]{\label{fig:restults:meta-frame-annotation}\includegraphics[width=0.2\textwidth]{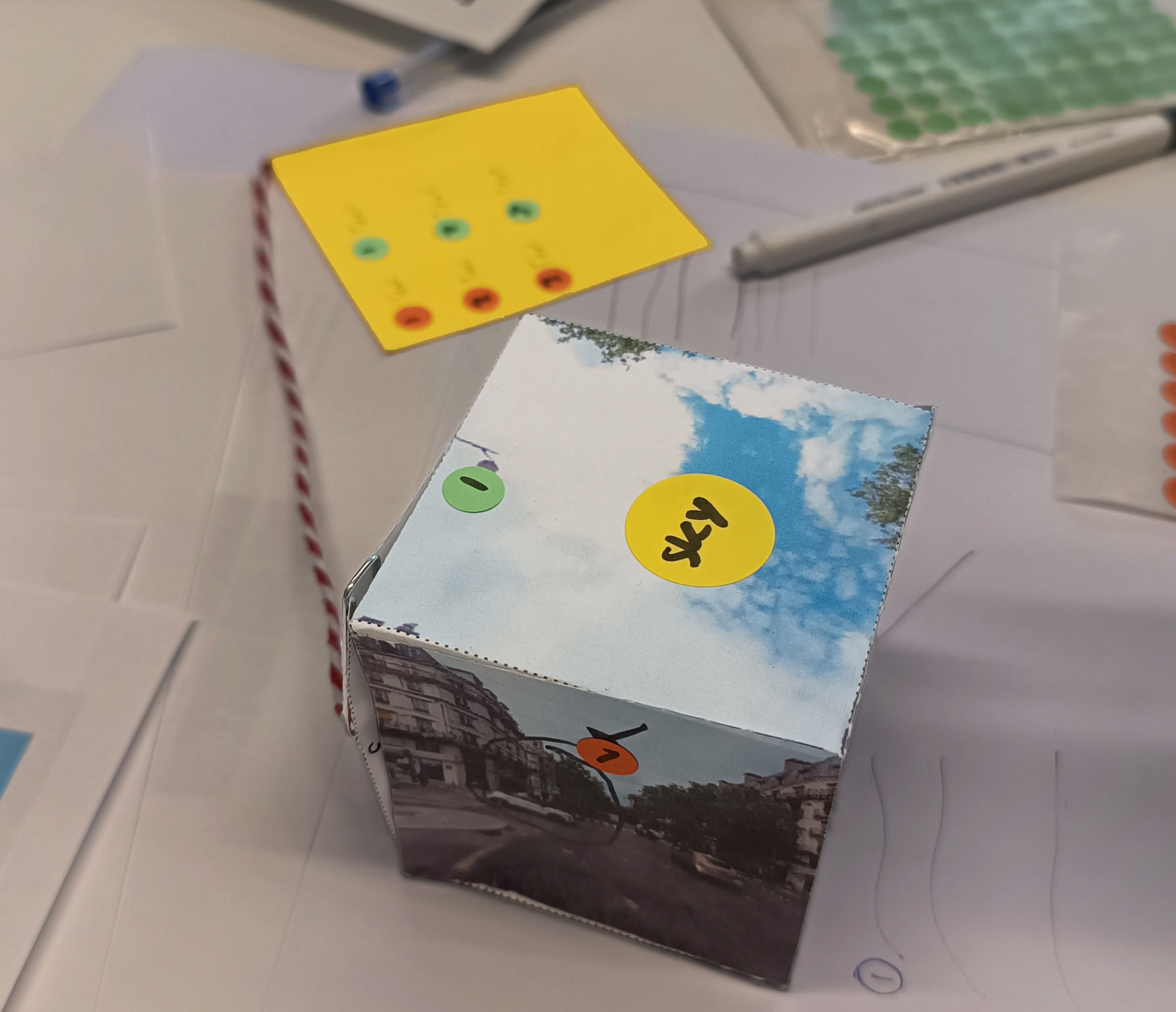}\Description{the cube shown in figure 9 b with some circular stickers attached to it to indicate different insights. There is a sticky note attached to the cube on a string which has text corresponding to the stickers.}}
    \qquad
    \subfloat[An example of combining the motion over several frames\protect\footnotemark{} in one 360° image suggested by P2, P3, P4, and P9.]{\label{fig:restults:motion-frames}\includegraphics[width=0.3\textwidth]{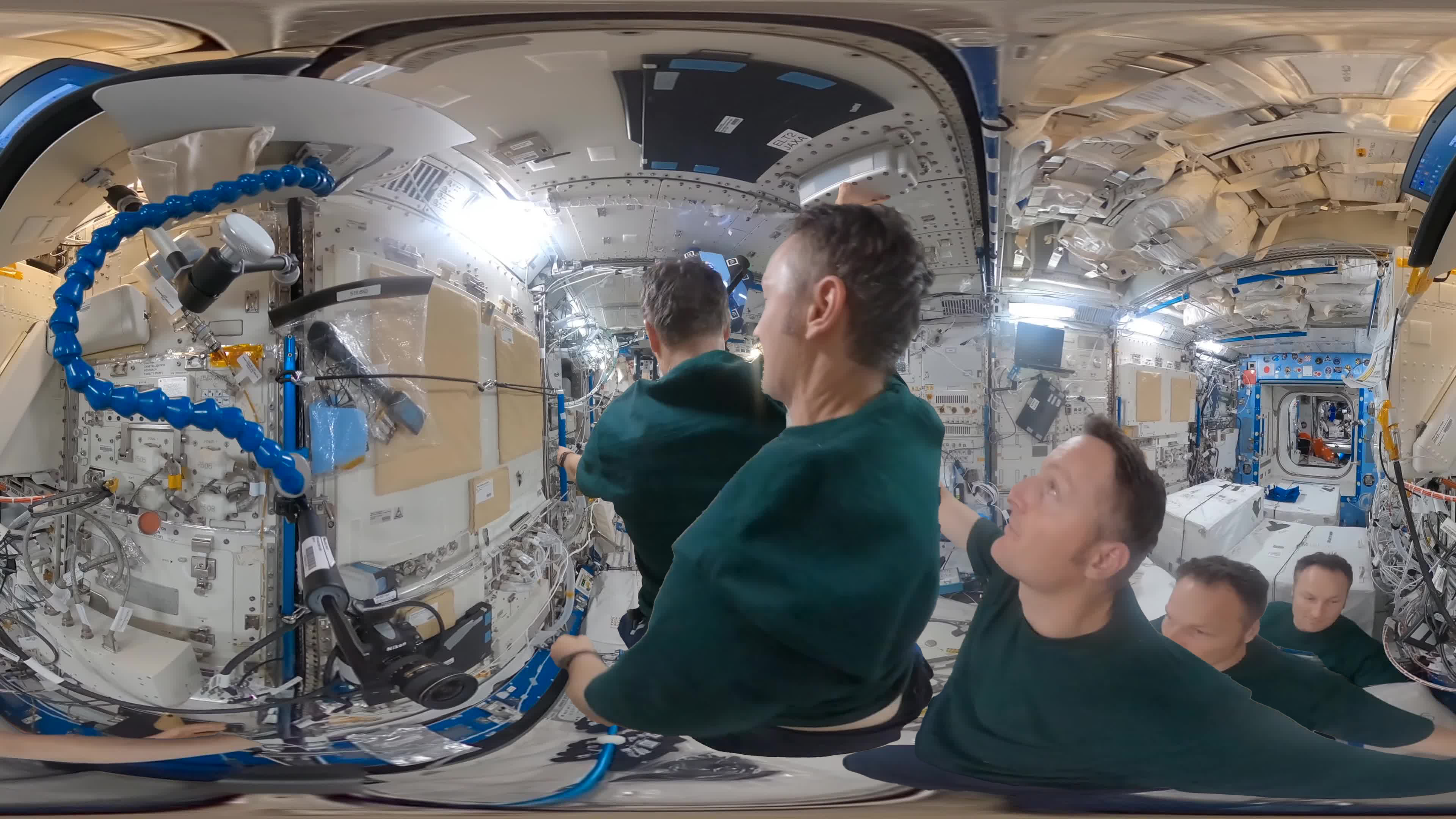}\Description{a single frame of 360° video that has multiple copies of the same person on it, indicating their motion around the frame}}
    \qquad
    \subfloat[A recreation of the multi-cube artifact suggested by P8 that maps part of a bicycle journey where the cyclist turns right.]{\label{fig:results:multi-cube}\includegraphics[width=0.3\textwidth]{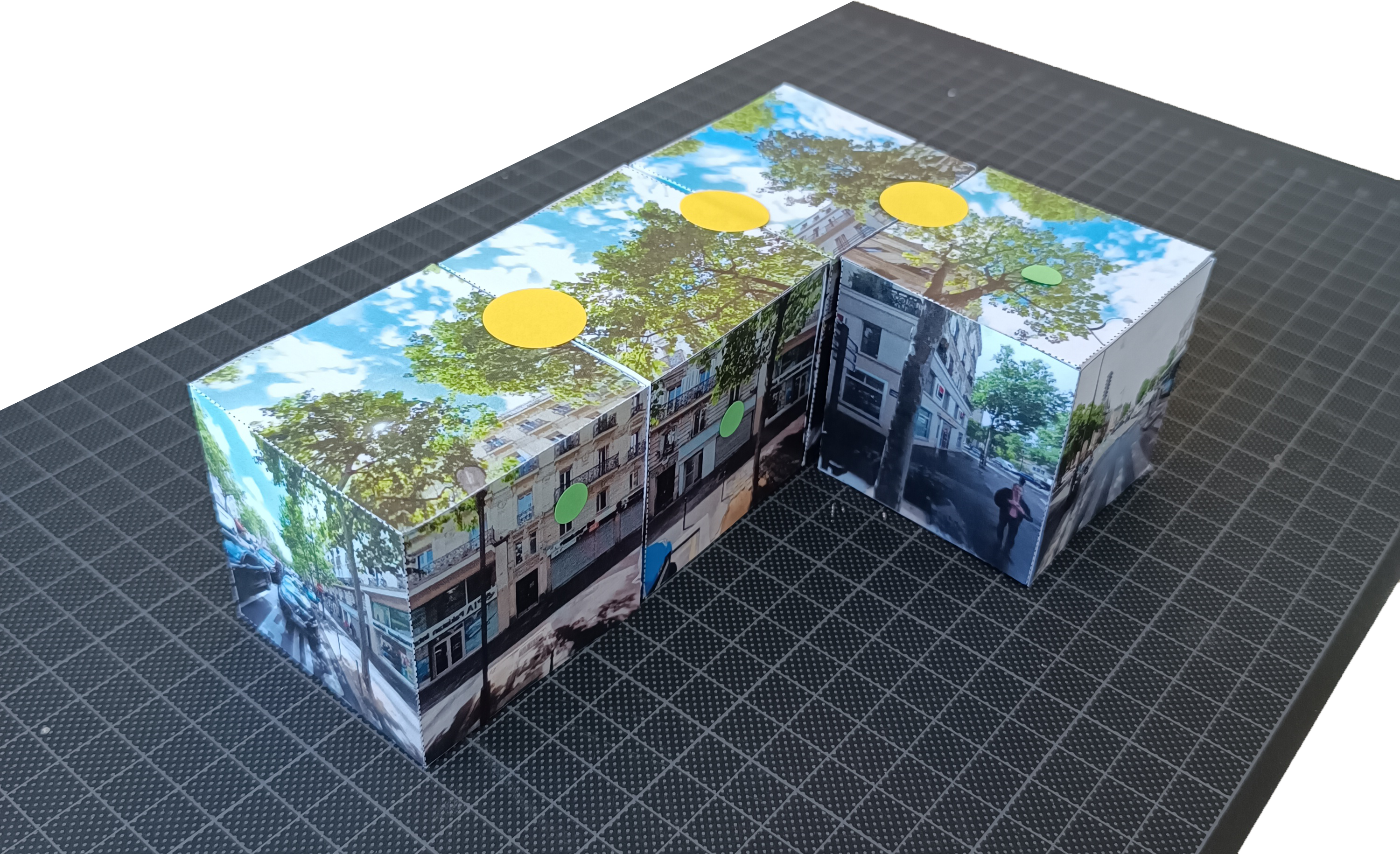}\Description{a series of paper cubes attached together, were each is a different frame of a video showing the motion of a cyclist.}}
    \caption{Additional examples of artifacts created by participants that go beyond the ``frame'' artifacts used as the basis of the sessions.}
    \label{fig:restults:beyond-frame-2}
\end{figure*}

\footnotetext{Footage from: Astrobee robots in 360° | Cosmic Kiss - European Space Agency - CC-BY. \url{https://youtu.be/ZfFssKBiOn8}}
\begin{enumerate}[label= M\arabic*]
    \item \label{result:sphere-meta-frame} \textbf{Sphere-ish meta-frame elements:} one implicit difference between the flat and sphere-ish artifacts is the sphere-ish artifacts do not have any white-space for meta-frame elements. P5, P6, P8, and P9 experimented with solutions to this issue, with both P6 and P9 suggesting an external \textit{``booklet''}~(P6) or \textit{``dog tags''}~(P9) joined to the artifact with string, which contain the detailed meta-annotation text or drawings connected to color-coded dots on the main artifact (see Figure~\ref{fig:restults:meta-frame-annotation}). P5 suggested the use of cut up sticky-notes used tangentially to the artifact, while P8 suggested using flags (e.g., small pieces of paper on sticks), magnetically attached notes, and QR code markers printed on stickers that would connect to an app. 
\end{enumerate}
\begin{enumerate}[resume,label= M\arabic*]
    \item \label{result:mutli-frame} \textbf{Multi-class artifacts:} One suggestion by P5 and P8 was the creation of multi-frame artifacts that combined the flat and sphere-ish artifacts to leverage the details of the flat artifact with the overview of the sphere-ish ones (see~\ref{result:difference-flat-sphere-ish}). Notably, P5 suggested using the sphere-ish artifact as the main artifact, with the flat artifacts either attached with strings where necessary or attached to the main artifact and unfolding like origami or a pop-up book. P9 on the other hand suggested the using all the artifacts like a time-line, where the choice of using a flat or sphere-ish artifact depending on the amount of detail or context deemed important at that moment.
\end{enumerate}
\begin{enumerate}[resume,label= M\arabic*]
    \item \label{result:miscellaneous} \textbf{Video specific} ideas: the artifacts that Tangi produces can support the use of both 360° photos and videos, participants came up with suggestions for novel artifacts that are specifically video based.
    \begin{enumerate}
        \item \textbf{Motion analysis:} P2, P3, P4, and P9 all suggested that, when the background of the video remains static, it is possible to show the motion of certain actors within a video by overlapping multiple frames on one single sphere-ish artifact (an example is shown in Figure~\ref{fig:restults:motion-frames}).
        \end{enumerate}
\begin{enumerate}[resume]
        \item \textbf{Timeline artifacts:} P8 suggested creating artifacts that join multiple frames of 360° video together in order to show not only a single moment in a 360° video, but to act as a multi-frame artifact (i.e., a timeline). This was roughly mocked-up by taping cube artifacts together, as shown in Figure~\ref{fig:results:multi-cube}.
    \end{enumerate}
\end{enumerate}

\noindent\textbf{Tangibility and Time}
Two observations made during the sessions by the researcher was how the intrinsic properties of the tangible artifacts impacted the sessions, specifically when creating or modifying the example artifacts and frustrations that surfaced during the sessions.
\begin{enumerate}[label= T\arabic*]
    \item \label{result:paper-based-benefits} \textbf{Ease of modifying paper:} because of the paper-based nature of the tangible artifacts used in this study, participants were able to easily cut, fold, and draw on the example artifacts to create their intended changes. P5, P6, P7, and P8 all used multiple print-outs of the same artifact to experiment with creating the artifacts described above. P6 was particularly enthusiastic when the lead researcher gave them permission to cut open one of the example artifacts, since they were simple to replace. Similarly, P8 used a pen to indicate how their eyes moved around the sphere-ish artifacts after the researcher reminded them the artifacts were simply paper.
\end{enumerate}
\begin{enumerate}[resume,label= T\arabic*]
    \item \label{result:fragility} \textbf{Fragility and folding time:} while the paper nature of the artifacts made them easy to modify, they also made them fragile. P1, P3, P6, and P9 all broke one of the pre-assembled artifacts during their respective sessions. While this was not a costly mistake; cutting, gluing, and folding a sphere-ish artifact was time-consuming process, especially during the time constrained sessions. For this reason, P2, P3, and P4 only worked with pre-assembled artifacts. Besides the restrictions brought upon by time, the artifacts were also fragile, consisting of printed paper and simple glue. For example, P9 attempted to hastily cut and fold an octahedron, but ended up gluing it incorrectly which lead to frustration and P9 abandoning the attempt at creating the example artifact.
\end{enumerate}
\section{Discussion}\label{discussion}
Our analysis demonstrates how participants were able to use the example tangible 360° video artifacts shared during the sessions as artifacts to both analyze 360° video and share the resulting insights -- a crucial step in enabling 360° VDE.
The contextually rich nature of the sphere-ish artifacts lend themselves well to immersion and familiarization, while the flat artifacts gave a more focused view for sharing specific insights (\ref{result:difference-flat-sphere-ish}). 
The study also elicited two uses of tangible artifacts specific to 360° video: supporting viewer orientation in the 360° space and linking the overall context to specific details.
Additionally, the tangible nature of the artifacts enabled participants to prototype a series of more complex artifacts, demonstrating the ability of paper-based 360° artifacts to support the evolving needs of a design team~\cite{dalsgaard_emergent_2014}/
In this section we will discuss how these insights can support a wider use of 360° video in design and beyond design, how Tangi will be expanded to support this, and finally the limitations of this study.

\subsection{Paper Based Artifacts Support Bricolage}\label{dis:bricolage}
Participants expressed a wide range of preferences for which of the artifacts Tangi creates in different stages of the design process (see~\ref{result:difference-flat-sphere-ish} and ~\ref{result:difference-cube-octaherdon}). 
The combination of personal preferences and purpose of the tangible artifacts reflects the concept of emergent boundary objects~\cite{dalsgaard_emergent_2014}. 
During the sessions, the ease of modifying the paper-based artifacts (\ref{result:paper-based-benefits}) let participants to quickly prototype changes or even novel artifacts. 
This enabled them to explore different ways to experience and document the 360° video without needing to switch from engaging in the physical workshop to a digital tool.
By giving participants this ability to tangibly interact with an manipulate 360° material (both photos and frames of video), the artifacts enabled them to engage in ``bricolage''~\cite{louridas_design_1999}, using the material at hand and its implicit restrictions to generate new artifacts and new insights while using them.
For example, designers could cut out important actors from a series of artifacts and overlay them to quickly create an artifact that demonstrates the motion in a scene (similar to the concept shown in Figure~\ref{fig:restults:motion-frames}).
Giving designers the ability to tangibly interact with and modify modify 360° video opens up new avenues for creativity and collaborative analysis through making without breaking up workshops with digital tools~\cite{buur_ethnographic_2010}.
Finally, designers are ``forced'' to reengage with the moments of 360° video when creating and modifying Tangi's artifacts which evolves the artifacts beyond a simple boundary object and become part of the analysis and reflection process.

\subsection{Tangi Beyond VDE}\label{dis:tangi-beyond-vde}
This paper focuses specifically on the use of (360°) video by designers for user research -- however, there are other ways designers use video and other users of video.
\citet{buur_video_2000} describes two ways in which video is used in design processes: in early stages, teams work on sense-making (i.e., the focus of this paper), but in later stages of design processes, design teams shift to using video as an evaluation material for prototypes and initial ideas.~\citet{ylirisku_designing_2007} goes further and discusses how designers use videos to make provocations in order to frame discussions with clients and the public about implications of future design ideas. These uses of video in design could also benefit from 360° video -- for example, by creating more immersive and complex \textit{360°} video provocations. 
These uses of 360° video as the output of a design process faces the same challenge of needing 360° video artifacts to share insights and frame discussions that are discussed in this paper. Based on the utility of the tangible frame artifacts Tangi generates to support the creation of more complex artifacts (\ref{result:mutli-frame}), designers can leverage Tangi to quickly and iteratively create bespoke types of tangible artifacts for provocations. Therefore, Tangi enables the exploration and study of how designers can use 360° video for prototypes and provocations with clients or the public as a whole.

Additionally, there are many uses of 360° video outside of design; such as education~\cite{snelson_educational_2020}, understanding urban environments~\cite{kim_capturing_2022}, or immersive journalism~\cite{mabrook_virtual_2019}. A specific example is the research of~\citet{sarkar_360-degree_2022} on how 360° video supports firefighter training --enabling instructors to illustrate important actions in an environment that is difficult to recreate (i.e., burning buildings). In this context, the firefighter instructor needs to share insights (objects and events that require an alert) that happen in a wide visual context (necessitating 360° video) with a group with different experiences (students). This process of sharing and discussing insights mirrors the activity of designers in VDE~\cite{buur_ethnographic_2010}, and thus the artifacts created by Tangi could help support these kinds of 360° video-based instruction sessions for firefighters~\cite{sarkar_360-degree_2022} or medical personnel~\cite{petrica_using_2021, davidsen_360vr_2022}.
 

\subsection{Future Work for Tangi}
Both the participants and the authors of this paper found several opportunities to improve the current tool and the artifacts it produces (see~\ref{result:fragility}). Based on this feedback, we have open-sourced the tool ([\url{removed-for-annonimity}]) and are working on:

\noindent \textbf{Improvements to the software}, by providing the possibility to view a 360° video and take screenshots within Tangi itself.
Currently, designers are required to have their own 360° video viewing software, adding to the complexity of adopting 360° VDE. By creating a full web based interface, this new version of Tangi would lower the threshold for designers working with 360° video even more.


\noindent \textbf{Improve durability of artifacts} by creating a 3D printed ``core'' to which the cut and fold patterns get attached by double-sided tape or using glue -- also removing the need to cut out the flaps for glue. This was suggested by P7 when they witnessed the researcher struggling to glue a cuboctahedron.
These strengthened artifacts would not only improve their persistence but also allow for more freedom in interaction (e.g., rolling or throwing artifacts) during discussions.
These improvements would also make installations where the public engages with 360° video artifacts more resilient.


\subsubsection{From Tangible Interaction to Tangible User Interface}
P7's suggestion to use QR code stickers for meta-annotations (see~\ref{result:sphere-meta-frame}) opened up the concept of using the tangible artifacts as tokens not just for the concept of a video at a specific moment~\cite{markopoulos_designing_2016}, but also as a token for a tangible user interface for video~\cite{zigelbaum_tangible_2007}.
This bridge to digital technologies could connect static artifacts with an important element of videos -- time.
For example, artifacts could act as bookmark for specific 360° videos at specific orientations, allowing designers to quickly re-engage with the 360° video that lead to the insight.
Tracking artifacts could enable designers to use augmented reality to view these videos or to overlay dynamic meta-annotations, leading to more complex interactions without breaking the tangible interactions, reducing the burden of digital interactions breaking collaboration~\cite{buur_taking_2000}.
While adding a QR code to the artifacts would obscure some visual information, this could be avoided with techniques such as steganography~\cite{tancik_stegastamp_2020} or infrared QR codes~\cite{dogan_standarone_2023} to embed a marker into the facets of the artifact without impacting the visual information.

\subsubsection{Enabling a Community Approach}
Our evaluation of Tangi and the artifacts it creates demonstrates the value and more importantly the flexibilty of paper-based artifacts for 360° video (and photos for that matter).
However, there are many types of (360°) video artifacts -- with value that depends on the goals of the designers and the stage of the design process~{\cite{dalsgaard_emergent_2014}}.
Investigating how Tangi (and the themes described in this paper) support diverse (and highly contextual) design work, requires giving as many designers as possible access to Tangi.
Therefor, we are expanding the Tangi website to support a community of practitioners to share how the artifacts they created as case studies.
This will support both the development of a taxonomy of different artifacts and wide-spread experimentation with tangible 360° video artifacts -- leading to new avenues for both design practice and research.
By engaging with both the academic community (through this paper) and with practicing designers (using the Tangi tool and online community), the impact of tangible 360° video artifacts can evolve beyond this initial work.

\subsection{Limitations}
We were only able to engage with a limited number of designers experienced with 360° video (largely due to its novelty~\cite{tojo_how_2021}) for a single session.

While this session did show the \textit{utility} of Tangi to support collaborative design ethnography sessions, it is difficult to demonstrate how \textit{effective} Tangi is for all designers -- especially if viewed with the paradigm that all design problems are essentially unique~{\cite{dorst_comparing_1995}}.
Future work should evaluate Tangi in a number of different design contexts with a focus on how tangible 360° video artifacts evolve{~\cite{dalsgaard_emergent_2014}} over time.

We aim to support this by open-sourcing Tangi, thereby enabling other designers and researchers to create artifacts, explore how and when they are useful, and even generate new types of artifacts, and thus knowledge, based on this work.

Finally, Tangi is an initial approach to creating tangible artifacts for 360° video, as such we limited the scope of our tool to a few artifacts, while there are limitless approaches for different flat and sphere-ish artifacts. Additionally, we only explored a single scale of sphere-ish artifact based on the maximum sized artifact created using standard A4 paper. While this helped keep workshops focused on the overall utility of the tangible 360° video artifacts, there are countless possible variations. By providing Tangi as an open-source tool, we aim to enable a diverse community that can explore and evaluate a variety of shapes, projections, materials, and types of artifact.
~\citet{he_data_2024}, for example, discuss how the scale and material choice of tangible cubes impact the interaction. Since there are no technical limitations to the size of Tangi's artifacts\footnote{Outside of scaling artifacts (not the ones discussed in this study) and practical issues finding large enough paper.}, future studies can use Tangi to explore how the impact of scale and material choice influence the utility and experience of interacting with 360° video.
\section{Conclusion}\label{conclusion}
In this paper we discuss the importance of creating tangible 360° video artifacts to support 360° Video Design Ethnography.
To provide designers with such tangible artifacts, we developed Tangi, an online tool designed to swiftly create two distinct types of tangible 360° frame artifacts: flat and sphere-like.
Using Tangi and example artifacts generated by it, we conducted nine sessions with designers experienced in 360° video which demonstrated the utility of these artifacts to support the interactions found in conventional VDE.
Participants were enthusiastic about the utility of tangible 360° video artifacts to support collaborative design work, and our analysis elicited two new functions of video artifacts: orientation and linking context with details.
Additionally, participants created several novel artifacts using Tangi, demonstrating the value of paper-based artifacts to enable the creation of emergent boundary objects~\cite{dalsgaard_emergent_2014}.

By offering a straightforward, accessible, and open-source tool, our goal is to enable a broad range of designers and researchers to engage with 360° video to develop deeper insights and consequently, better solutions tailored to a diverse array of users and contexts. Additionally, Tangi can support sharing 360° video insights beyond the initial design process (e.g., evaluation and provocation) and beyond design (e.g., education and training).
\bibliographystyle{ACM-Reference-Format}
\bibliography{references}


\begin{thebibliography}{53}


\ifx \showCODEN    \undefined \def \showCODEN     #1{\unskip}     \fi
\ifx \showDOI      \undefined \def \showDOI       #1{#1}\fi
\ifx \showISBNx    \undefined \def \showISBNx     #1{\unskip}     \fi
\ifx \showISBNxiii \undefined \def \showISBNxiii  #1{\unskip}     \fi
\ifx \showISSN     \undefined \def \showISSN      #1{\unskip}     \fi
\ifx \showLCCN     \undefined \def \showLCCN      #1{\unskip}     \fi
\ifx \shownote     \undefined \def \shownote      #1{#1}          \fi
\ifx \showarticletitle \undefined \def \showarticletitle #1{#1}   \fi
\ifx \showURL      \undefined \def \showURL       {\relax}        \fi
\providecommand\bibfield[2]{#2}
\providecommand\bibinfo[2]{#2}
\providecommand\natexlab[1]{#1}
\providecommand\showeprint[2][]{arXiv:#2}

\bibitem[~(2017)]%
        {noauthor_saya360_2017}
\bibfield{author}{\bibinfo{person}{sizima~soft  }.} \bibinfo{year}{2017}\natexlab{}.
\newblock \bibinfo{title}{{SAYA360}}.
\newblock
\newblock
\urldef\tempurl%
\url{https://sizima.com/paper360/index_jp.html}
\showURL{%
\tempurl}


\bibitem[Beyer and Holtzblatt(1999)]%
        {beyer_contextual_1999}
\bibfield{author}{\bibinfo{person}{Hugh Beyer} {and} \bibinfo{person}{Karen Holtzblatt}.} \bibinfo{year}{1999}\natexlab{}.
\newblock \showarticletitle{Contextual design}.
\newblock \bibinfo{journal}{\emph{Interactions}} \bibinfo{volume}{6}, \bibinfo{number}{1} (\bibinfo{date}{Jan.} \bibinfo{year}{1999}), \bibinfo{pages}{32--42}.
\newblock
\showISSN{1072-5520}
\urldef\tempurl%
\url{https://doi.org/10.1145/291224.291229}
\showDOI{\tempurl}


\bibitem[Brandt(2007)]%
        {brandt_how_2007}
\bibfield{author}{\bibinfo{person}{Eva Brandt}.} \bibinfo{year}{2007}\natexlab{}.
\newblock \showarticletitle{How {Tangible} {Mock}-{Ups} {Support} {Design} {Collaboration}}.
\newblock \bibinfo{journal}{\emph{Knowledge, Technology \& Policy}} \bibinfo{volume}{20}, \bibinfo{number}{3} (\bibinfo{date}{Oct.} \bibinfo{year}{2007}), \bibinfo{pages}{179--192}.
\newblock
\showISSN{1874-6314}
\urldef\tempurl%
\url{https://doi.org/10.1007/s12130-007-9021-9}
\showDOI{\tempurl}


\bibitem[Braun and Clarke(2006)]%
        {braun_using_2006}
\bibfield{author}{\bibinfo{person}{Virginia Braun} {and} \bibinfo{person}{Victoria Clarke}.} \bibinfo{year}{2006}\natexlab{}.
\newblock \showarticletitle{Using thematic analysis in psychology}.
\newblock \bibinfo{journal}{\emph{Qualitative Research in Psychology}} \bibinfo{volume}{3}, \bibinfo{number}{2} (\bibinfo{date}{Jan.} \bibinfo{year}{2006}), \bibinfo{pages}{77--101}.
\newblock
\showISSN{1478-0887, 1478-0895}
\urldef\tempurl%
\url{https://doi.org/10.1191/1478088706qp063oa}
\showDOI{\tempurl}


\bibitem[Braun and Clarke(2022)]%
        {braun_thematic_2022}
\bibfield{author}{\bibinfo{person}{Virginia Braun} {and} \bibinfo{person}{Victoria Clarke}.} \bibinfo{year}{2022}\natexlab{}.
\newblock \bibinfo{booktitle}{\emph{Thematic analysis: a practical guide}}.
\newblock \bibinfo{publisher}{SAGE}, \bibinfo{address}{London ; Thousand Oaks, California}.
\newblock
\showISBNx{978-1-4739-5323-9 978-1-4739-5324-6}
\newblock
\shownote{OCLC: on1247204005}.


\bibitem[Buur et~al\mbox{.}(2000)]%
        {buur_taking_2000}
\bibfield{author}{\bibinfo{person}{Jacob Buur}, \bibinfo{person}{Thomas Binder}, {and} \bibinfo{person}{Eva Brandt}.} \bibinfo{year}{2000}\natexlab{}.
\newblock \showarticletitle{Taking {Video} {Beyond} '{Hard} {Data}' in {User} {Centered} {Design}}. In \bibinfo{booktitle}{\emph{In {Proceedings} of {Participatory} {Design} {Conference}}}. \bibinfo{pages}{21--29}.
\newblock


\bibitem[Buur et~al\mbox{.}(2010)]%
        {buur_ethnographic_2010}
\bibfield{author}{\bibinfo{person}{Jacob Buur}, \bibinfo{person}{Euan Fraser}, \bibinfo{person}{Soila Oinonen}, {and} \bibinfo{person}{Max Rolfstam}.} \bibinfo{year}{2010}\natexlab{}.
\newblock \showarticletitle{Ethnographic video as design specs}. In \bibinfo{booktitle}{\emph{Proceedings of the 22nd {Conference} of the {Computer}-{Human} {Interaction} {Special} {Interest} {Group} of {Australia} on {Computer}-{Human} {Interaction}}} \emph{(\bibinfo{series}{{OZCHI} '10})}. \bibinfo{publisher}{Association for Computing Machinery}, \bibinfo{address}{New York, NY, USA}, \bibinfo{pages}{49--56}.
\newblock
\showISBNx{978-1-4503-0502-0}
\urldef\tempurl%
\url{https://doi.org/10.1145/1952222.1952235}
\showDOI{\tempurl}


\bibitem[Buur and Soendergaard(2000)]%
        {buur_video_2000}
\bibfield{author}{\bibinfo{person}{Jacob Buur} {and} \bibinfo{person}{Astrid Soendergaard}.} \bibinfo{year}{2000}\natexlab{}.
\newblock \showarticletitle{Video card game: an augmented environment for user centred design discussions}. In \bibinfo{booktitle}{\emph{Proceedings of {DARE} 2000 on {Designing} augmented reality environments}} \emph{(\bibinfo{series}{{DARE} '00})}. \bibinfo{publisher}{Association for Computing Machinery}, \bibinfo{address}{New York, NY, USA}, \bibinfo{pages}{63--69}.
\newblock
\showISBNx{978-1-4503-7326-5}
\urldef\tempurl%
\url{https://doi.org/10.1145/354666.354673}
\showDOI{\tempurl}


\bibitem[Dalsgaard et~al\mbox{.}(2014)]%
        {dalsgaard_emergent_2014}
\bibfield{author}{\bibinfo{person}{Peter Dalsgaard}, \bibinfo{person}{Kim Halskov}, {and} \bibinfo{person}{Ditte~Amund Basballe}.} \bibinfo{year}{2014}\natexlab{}.
\newblock \showarticletitle{Emergent boundary objects and boundary zones in collaborative design research projects}. In \bibinfo{booktitle}{\emph{Proceedings of the 2014 conference on {Designing} interactive systems}}. \bibinfo{publisher}{ACM}, \bibinfo{address}{Vancouver BC Canada}, \bibinfo{pages}{745--754}.
\newblock
\showISBNx{978-1-4503-2902-6}
\urldef\tempurl%
\url{https://doi.org/10.1145/2598510.2600878}
\showDOI{\tempurl}


\bibitem[Davidsen et~al\mbox{.}(2022)]%
        {davidsen_360vr_2022}
\bibfield{author}{\bibinfo{person}{Jacob Davidsen}, \bibinfo{person}{Dorthe~Vinther Larsen}, \bibinfo{person}{Lucas Paulsen}, {and} \bibinfo{person}{Sten Rasmussen}.} \bibinfo{year}{2022}\natexlab{}.
\newblock \showarticletitle{{360VR} {PBL}: {A} {New} {Format} of {Digital} {Cases} in {Clinical} {Medicine}}.
\newblock \bibinfo{journal}{\emph{Journal of Problem Based Learning in Higher Education}} \bibinfo{volume}{10}, \bibinfo{number}{1} (\bibinfo{date}{Oct.} \bibinfo{year}{2022}).
\newblock
\showISSN{2246-0918}
\urldef\tempurl%
\url{https://doi.org/10.54337/ojs.jpblhe.v10i1.7097}
\showDOI{\tempurl}
\newblock
\shownote{Number: 1}.


\bibitem[Dogan et~al\mbox{.}(2023)]%
        {dogan_standarone_2023}
\bibfield{author}{\bibinfo{person}{Mustafa~Doga Dogan}, \bibinfo{person}{Alexa~F. Siu}, \bibinfo{person}{Jennifer Healey}, \bibinfo{person}{Curtis Wigington}, \bibinfo{person}{Chang Xiao}, {and} \bibinfo{person}{Tong Sun}.} \bibinfo{year}{2023}\natexlab{}.
\newblock \showarticletitle{{StandARone}: {Infrared}-{Watermarked} {Documents} as {Portable} {Containers} of {AR} {Interaction} and {Personalization}}. In \bibinfo{booktitle}{\emph{Extended {Abstracts} of the 2023 {CHI} {Conference} on {Human} {Factors} in {Computing} {Systems}}} \emph{(\bibinfo{series}{{CHI} {EA} '23})}. \bibinfo{publisher}{Association for Computing Machinery}, \bibinfo{address}{New York, NY, USA}, \bibinfo{pages}{1--7}.
\newblock
\showISBNx{978-1-4503-9422-2}
\urldef\tempurl%
\url{https://doi.org/10.1145/3544549.3585905}
\showDOI{\tempurl}


\bibitem[Donen(1963)]%
        {donen_charade_1963}
\bibfield{author}{\bibinfo{person}{Stanley Donen}.} \bibinfo{year}{1963}\natexlab{}.
\newblock \bibinfo{title}{Charade}.
\newblock
\newblock


\bibitem[Dorst and Dijkhuis(1995)]%
        {dorst_comparing_1995}
\bibfield{author}{\bibinfo{person}{Kees Dorst} {and} \bibinfo{person}{Judith Dijkhuis}.} \bibinfo{year}{1995}\natexlab{}.
\newblock \showarticletitle{Comparing paradigms for describing design activity}.
\newblock \bibinfo{journal}{\emph{Design Studies}} \bibinfo{volume}{16}, \bibinfo{number}{2} (\bibinfo{date}{April} \bibinfo{year}{1995}), \bibinfo{pages}{261--274}.
\newblock
\showISSN{0142-694X}
\urldef\tempurl%
\url{https://doi.org/10.1016/0142-694X(94)00012-3}
\showDOI{\tempurl}


\bibitem[Goldman et~al\mbox{.}(2006)]%
        {goldman_schematic_2006}
\bibfield{author}{\bibinfo{person}{Dan~B Goldman}, \bibinfo{person}{Brian Curless}, \bibinfo{person}{David Salesin}, {and} \bibinfo{person}{Steven~M. Seitz}.} \bibinfo{year}{2006}\natexlab{}.
\newblock \showarticletitle{Schematic storyboarding for video visualization and editing}.
\newblock \bibinfo{journal}{\emph{ACM Transactions on Graphics}} \bibinfo{volume}{25}, \bibinfo{number}{3} (\bibinfo{date}{July} \bibinfo{year}{2006}), \bibinfo{pages}{862--871}.
\newblock
\showISSN{0730-0301, 1557-7368}
\urldef\tempurl%
\url{https://doi.org/10.1145/1141911.1141967}
\showDOI{\tempurl}


\bibitem[He et~al\mbox{.}(2024)]%
        {he_data_2024}
\bibfield{author}{\bibinfo{person}{Shuqi He}, \bibinfo{person}{Haonan Yao}, \bibinfo{person}{Luyan Jiang}, \bibinfo{person}{Kaiwen Li}, \bibinfo{person}{Nan Xiang}, \bibinfo{person}{Yue Li}, \bibinfo{person}{Hai-Ning Liang}, {and} \bibinfo{person}{Lingyun Yu}.} \bibinfo{year}{2024}\natexlab{}.
\newblock \showarticletitle{Data {Cubes} in {Hand}: {A} {Design} {Space} of {Tangible} {Cubes} for {Visualizing} {3D} {Spatio}-{Temporal} {Data} in {Mixed} {Reality}}. In \bibinfo{booktitle}{\emph{Proceedings of the {CHI} {Conference} on {Human} {Factors} in {Computing} {Systems}}} \emph{(\bibinfo{series}{{CHI} '24})}. \bibinfo{publisher}{Association for Computing Machinery}, \bibinfo{address}{New York, NY, USA}, \bibinfo{pages}{1--21}.
\newblock
\showISBNx{9798400703300}
\urldef\tempurl%
\url{https://doi.org/10.1145/3613904.3642740}
\showDOI{\tempurl}


\bibitem[Hurbain(2003)]%
        {hurbain_3-d_2003}
\bibfield{author}{\bibinfo{person}{Philippe Hurbain}.} \bibinfo{year}{2003}\natexlab{}.
\newblock \bibinfo{title}{3-{D} panorama printing: enter physical reality...}
\newblock
\newblock
\urldef\tempurl%
\url{https://web.archive.org/web/20030413190851/https://www.philohome.com/rhombicuboctahedron/rhombicuboctahedron.htm}
\showURL{%
\tempurl}


\bibitem[Jokela et~al\mbox{.}(2019)]%
        {jokela_how_2019}
\bibfield{author}{\bibinfo{person}{Tero Jokela}, \bibinfo{person}{Jarno Ojala}, {and} \bibinfo{person}{Kaisa Väänänen}.} \bibinfo{year}{2019}\natexlab{}.
\newblock \showarticletitle{How people use 360-degree cameras}. In \bibinfo{booktitle}{\emph{Proceedings of the 18th {International} {Conference} on {Mobile} and {Ubiquitous} {Multimedia}}}. \bibinfo{publisher}{ACM}, \bibinfo{address}{Pisa Italy}, \bibinfo{pages}{1--10}.
\newblock
\showISBNx{978-1-4503-7624-2}
\urldef\tempurl%
\url{https://doi.org/10.1145/3365610.3365645}
\showDOI{\tempurl}


\bibitem[Kim and Lee(2022)]%
        {kim_capturing_2022}
\bibfield{author}{\bibinfo{person}{Seung-Nam Kim} {and} \bibinfo{person}{Hanwool Lee}.} \bibinfo{year}{2022}\natexlab{}.
\newblock \showarticletitle{Capturing reality: {Validation} of omnidirectional video-based immersive virtual reality as a streetscape quality auditing method}.
\newblock \bibinfo{journal}{\emph{Landscape and Urban Planning}}  \bibinfo{volume}{218} (\bibinfo{date}{Feb.} \bibinfo{year}{2022}), \bibinfo{pages}{104290}.
\newblock
\showISSN{0169-2046}
\urldef\tempurl%
\url{https://doi.org/10.1016/j.landurbplan.2021.104290}
\showDOI{\tempurl}


\bibitem[Koenderink and van Doorn(2017)]%
        {koenderink_planispheric_2017}
\bibfield{author}{\bibinfo{person}{Jan Koenderink} {and} \bibinfo{person}{Andrea van Doorn}.} \bibinfo{year}{2017}\natexlab{}.
\newblock \showarticletitle{The {Planispheric} {Optic} {Array}}.
\newblock \bibinfo{journal}{\emph{i-Perception}} \bibinfo{volume}{8}, \bibinfo{number}{3} (\bibinfo{date}{June} \bibinfo{year}{2017}), \bibinfo{pages}{2041669517705388}.
\newblock
\showISSN{2041-6695}
\urldef\tempurl%
\url{https://doi.org/10.1177/2041669517705388}
\showDOI{\tempurl}
\newblock
\shownote{Publisher: SAGE Publications}.


\bibitem[Kosko et~al\mbox{.}(2022)]%
        {kosko_using_2022}
\bibfield{author}{\bibinfo{person}{Karl~W. Kosko}, \bibinfo{person}{Jennifer Heisler}, {and} \bibinfo{person}{Enrico Gandolfi}.} \bibinfo{year}{2022}\natexlab{}.
\newblock \showarticletitle{Using 360-degree video to explore teachers' professional noticing}.
\newblock \bibinfo{journal}{\emph{Computers \& Education}}  \bibinfo{volume}{180} (\bibinfo{date}{April} \bibinfo{year}{2022}), \bibinfo{pages}{104443}.
\newblock
\showISSN{0360-1315}
\urldef\tempurl%
\url{https://doi.org/10.1016/j.compedu.2022.104443}
\showDOI{\tempurl}


\bibitem[Kramer(2022)]%
        {kramer_innovating_2022}
\bibfield{author}{\bibinfo{person}{Mark A.~M. Kramer}.} \bibinfo{year}{2022}\natexlab{}.
\newblock \showarticletitle{Innovating {Methods}: {Observing} {In} {Situ} {Technology} {Experiences} {Using} {Novel} {Visual} {Research} {Methods}}.
\newblock \bibinfo{journal}{\emph{International Journal of Human–Computer Interaction}} \bibinfo{volume}{38}, \bibinfo{number}{3} (\bibinfo{date}{Feb.} \bibinfo{year}{2022}), \bibinfo{pages}{253--262}.
\newblock
\showISSN{1044-7318}
\urldef\tempurl%
\url{https://doi.org/10.1080/10447318.2021.1938388}
\showDOI{\tempurl}
\newblock
\shownote{Publisher: Taylor \& Francis \_eprint: https://doi.org/10.1080/10447318.2021.1938388}.


\bibitem[Ledo et~al\mbox{.}(2018)]%
        {ledo_evaluation_2018}
\bibfield{author}{\bibinfo{person}{David Ledo}, \bibinfo{person}{Steven Houben}, \bibinfo{person}{Jo Vermeulen}, \bibinfo{person}{Nicolai Marquardt}, \bibinfo{person}{Lora Oehlberg}, {and} \bibinfo{person}{Saul Greenberg}.} \bibinfo{year}{2018}\natexlab{}.
\newblock \showarticletitle{Evaluation {Strategies} for {HCI} {Toolkit} {Research}}. In \bibinfo{booktitle}{\emph{Proceedings of the 2018 {CHI} {Conference} on {Human} {Factors} in {Computing} {Systems}}}. \bibinfo{publisher}{ACM}, \bibinfo{address}{Montreal QC Canada}, \bibinfo{pages}{1--17}.
\newblock
\showISBNx{978-1-4503-5620-6}
\urldef\tempurl%
\url{https://doi.org/10.1145/3173574.3173610}
\showDOI{\tempurl}


\bibitem[Lee(1944)]%
        {lee_nomenclature_1944}
\bibfield{author}{\bibinfo{person}{L~P Lee}.} \bibinfo{year}{1944}\natexlab{}.
\newblock \showarticletitle{The {Nomenclature} and {Classification} of {Map} {Projections}}.
\newblock \bibinfo{journal}{\emph{Empire Survey Review}} \bibinfo{volume}{7}, \bibinfo{number}{50} (\bibinfo{date}{Jan.} \bibinfo{year}{1944}), \bibinfo{pages}{190--200}.
\newblock


\bibitem[Li et~al\mbox{.}(2021)]%
        {li_route_2021}
\bibfield{author}{\bibinfo{person}{Jiannan Li}, \bibinfo{person}{Jiahe Lyu}, \bibinfo{person}{Mauricio Sousa}, \bibinfo{person}{Ravin Balakrishnan}, \bibinfo{person}{Anthony Tang}, {and} \bibinfo{person}{Tovi Grossman}.} \bibinfo{year}{2021}\natexlab{}.
\newblock \showarticletitle{Route {Tapestries}: {Navigating} 360° {Virtual} {Tour} {Videos} {Using} {Slit}-{Scan} {Visualizations}}. In \bibinfo{booktitle}{\emph{The 34th {Annual} {ACM} {Symposium} on {User} {Interface} {Software} and {Technology}}}. \bibinfo{publisher}{ACM}, \bibinfo{address}{Virtual Event USA}, \bibinfo{pages}{223--238}.
\newblock
\showISBNx{978-1-4503-8635-7}
\urldef\tempurl%
\url{https://doi.org/10.1145/3472749.3474746}
\showDOI{\tempurl}


\bibitem[Li et~al\mbox{.}(2016)]%
        {li_omnieyeball_2016}
\bibfield{author}{\bibinfo{person}{Zhengqing Li}, \bibinfo{person}{Shio Miyafuji}, \bibinfo{person}{Toshiki Sato}, {and} \bibinfo{person}{Hideki Koike}.} \bibinfo{year}{2016}\natexlab{}.
\newblock \showarticletitle{{OmniEyeball}: {Spherical} {Display} {Embedded} {With} {Omnidirectional} {Camera} {Using} {Dynamic} {Spherical} {Mapping}}. In \bibinfo{booktitle}{\emph{Adjunct {Proceedings} of the 29th {Annual} {ACM} {Symposium} on {User} {Interface} {Software} and {Technology}}} \emph{(\bibinfo{series}{{UIST} '16 {Adjunct}})}. \bibinfo{publisher}{Association for Computing Machinery}, \bibinfo{address}{New York, NY, USA}, \bibinfo{pages}{193--194}.
\newblock
\showISBNx{978-1-4503-4531-6}
\urldef\tempurl%
\url{https://doi.org/10.1145/2984751.2984765}
\showDOI{\tempurl}


\bibitem[Louridas(1999)]%
        {louridas_design_1999}
\bibfield{author}{\bibinfo{person}{Panagiotis Louridas}.} \bibinfo{year}{1999}\natexlab{}.
\newblock \showarticletitle{Design as bricolage: anthropology meets design thinking}.
\newblock \bibinfo{journal}{\emph{Design Studies}} \bibinfo{volume}{20}, \bibinfo{number}{6} (\bibinfo{date}{Nov.} \bibinfo{year}{1999}), \bibinfo{pages}{517--535}.
\newblock
\showISSN{0142-694X}
\urldef\tempurl%
\url{https://doi.org/10.1016/S0142-694X(98)00044-1}
\showDOI{\tempurl}


\bibitem[Lucero et~al\mbox{.}(2016)]%
        {markopoulos_designing_2016}
\bibfield{author}{\bibinfo{person}{Andrés Lucero}, \bibinfo{person}{Peter Dalsgaard}, \bibinfo{person}{Kim Halskov}, {and} \bibinfo{person}{Jacob Buur}.} \bibinfo{year}{2016}\natexlab{}.
\newblock \showarticletitle{Designing with {Cards}}.
\newblock In \bibinfo{booktitle}{\emph{Collaboration in {Creative} {Design}}}, \bibfield{editor}{\bibinfo{person}{Panos Markopoulos}, \bibinfo{person}{Jean-Bernard Martens}, \bibinfo{person}{Julian Malins}, \bibinfo{person}{Karin Coninx}, {and} \bibinfo{person}{Aggelos Liapis}} (Eds.). \bibinfo{publisher}{Springer International Publishing}, \bibinfo{address}{Cham}, \bibinfo{pages}{75--95}.
\newblock
\showISBNx{978-3-319-29153-6 978-3-319-29155-0}
\urldef\tempurl%
\url{https://doi.org/10.1007/978-3-319-29155-0_5}
\showDOI{\tempurl}


\bibitem[Mabrook and Singer(2019)]%
        {mabrook_virtual_2019}
\bibfield{author}{\bibinfo{person}{Radwa Mabrook} {and} \bibinfo{person}{Jane~B. Singer}.} \bibinfo{year}{2019}\natexlab{}.
\newblock \showarticletitle{Virtual {Reality}, 360° {Video}, and {Journalism} {Studies}: {Conceptual} {Approaches} to {Immersive} {Technologies}}.
\newblock \bibinfo{journal}{\emph{Journalism Studies}} \bibinfo{volume}{20}, \bibinfo{number}{14} (\bibinfo{date}{Oct.} \bibinfo{year}{2019}), \bibinfo{pages}{2096--2112}.
\newblock
\showISSN{1461-670X, 1469-9699}
\urldef\tempurl%
\url{https://doi.org/10.1080/1461670X.2019.1568203}
\showDOI{\tempurl}


\bibitem[MacEachren(2004)]%
        {maceachren_how_2004}
\bibfield{author}{\bibinfo{person}{Alan~M. MacEachren}.} \bibinfo{year}{2004}\natexlab{}.
\newblock \bibinfo{booktitle}{\emph{How maps work: representation, visualization, and design} (\bibinfo{edition}{paperback edition} ed.)}.
\newblock \bibinfo{publisher}{Guilford Press}, \bibinfo{address}{New York}.
\newblock
\showISBNx{978-1-57230-040-8}
\urldef\tempurl%
\url{http://catdir.loc.gov/catdir/enhancements/fy0626/2005276848-t.html}
\showURL{%
\tempurl}
\newblock
\shownote{OCLC: 56039088}.


\bibitem[Meijer et~al\mbox{.}(2024)]%
        {meijer_sphere_2024}
\bibfield{author}{\bibinfo{person}{Wo Meijer}, \bibinfo{person}{Jacky Bourgeois}, \bibinfo{person}{Wilhelm~Frederik van~der Vegte}, {and} \bibinfo{person}{Gerd Kortuem}.} \bibinfo{year}{2024}\natexlab{}.
\newblock \showarticletitle{Sphere {Window}: {Challenges} and {Opportunities} of 360° {Video} in {Collaborative} {Design} {Workshops}.}. In \bibinfo{booktitle}{\emph{Nordic {Conference} on {Human}-{Computer} {Interaction}}} \emph{(\bibinfo{series}{{NordiCHI} 2024})}. \bibinfo{publisher}{Association for Computing Machinery}, \bibinfo{address}{New York, NY, USA}.
\newblock
\urldef\tempurl%
\url{https://doi.org/10.1145/3679318.3685407}
\showDOI{\tempurl}


\bibitem[Neubauer et~al\mbox{.}(2017)]%
        {neubauer_experiencing_2017}
\bibfield{author}{\bibinfo{person}{Daniel Neubauer}, \bibinfo{person}{Verena Paepcke-Hjeltness}, \bibinfo{person}{Pete Evans}, \bibinfo{person}{Betsy Barnhart}, {and} \bibinfo{person}{Tor Finseth}.} \bibinfo{year}{2017}\natexlab{}.
\newblock \showarticletitle{Experiencing {Technology} {Enabled} {Empathy} {Mapping}}.
\newblock \bibinfo{journal}{\emph{The Design Journal}} \bibinfo{volume}{20}, \bibinfo{number}{sup1} (\bibinfo{date}{July} \bibinfo{year}{2017}), \bibinfo{pages}{S4683--S4689}.
\newblock
\showISSN{1460-6925}
\urldef\tempurl%
\url{https://doi.org/10.1080/14606925.2017.1352966}
\showDOI{\tempurl}
\newblock
\shownote{Publisher: Routledge \_eprint: https://doi.org/10.1080/14606925.2017.1352966}.


\bibitem[Nguyen et~al\mbox{.}(2017)]%
        {nguyen_vremiere_2017}
\bibfield{author}{\bibinfo{person}{Cuong Nguyen}, \bibinfo{person}{Stephen DiVerdi}, \bibinfo{person}{Aaron Hertzmann}, {and} \bibinfo{person}{Feng Liu}.} \bibinfo{year}{2017}\natexlab{}.
\newblock \showarticletitle{Vremiere: {In}-{Headset} {Virtual} {Reality} {Video} {Editing}}. In \bibinfo{booktitle}{\emph{Proceedings of the 2017 {CHI} {Conference} on {Human} {Factors} in {Computing} {Systems}}}. \bibinfo{publisher}{ACM}, \bibinfo{address}{Denver Colorado USA}, \bibinfo{pages}{5428--5438}.
\newblock
\showISBNx{978-1-4503-4655-9}
\urldef\tempurl%
\url{https://doi.org/10.1145/3025453.3025675}
\showDOI{\tempurl}


\bibitem[Nova(2014)]%
        {nova_beyond_2014}
\bibfield{author}{\bibinfo{person}{Nicolas Nova}.} \bibinfo{year}{2014}\natexlab{}.
\newblock \bibinfo{booktitle}{\emph{Beyond design ethnography}}.
\newblock \bibinfo{publisher}{SHS Publishing}.
\newblock
\showISBNx{978-88-907594-4-4}


\bibitem[Petrica et~al\mbox{.}(2021)]%
        {petrica_using_2021}
\bibfield{author}{\bibinfo{person}{Alina Petrica}, \bibinfo{person}{Diana Lungeanu}, \bibinfo{person}{Alexandru Ciuta}, \bibinfo{person}{Adina~M. Marza}, \bibinfo{person}{Mihai-Octavian Botea}, {and} \bibinfo{person}{Ovidiu~A. Mederle}.} \bibinfo{year}{2021}\natexlab{}.
\newblock \showarticletitle{Using 360-degree video for teaching emergency medicine during and beyond the {COVID}-19 pandemic}.
\newblock \bibinfo{journal}{\emph{Annals of Medicine}} \bibinfo{volume}{53}, \bibinfo{number}{1} (\bibinfo{date}{Jan.} \bibinfo{year}{2021}), \bibinfo{pages}{1520--1530}.
\newblock
\showISSN{0785-3890}
\urldef\tempurl%
\url{https://doi.org/10.1080/07853890.2021.1970219}
\showDOI{\tempurl}
\newblock
\shownote{Publisher: Taylor \& Francis \_eprint: https://doi.org/10.1080/07853890.2021.1970219}.


\bibitem[Pimentel et~al\mbox{.}(2021)]%
        {pimentel_voices_2021}
\bibfield{author}{\bibinfo{person}{Daniel Pimentel}, \bibinfo{person}{Sri Kalyanaraman}, \bibinfo{person}{Yu-Hao Lee}, {and} \bibinfo{person}{Shiva Halan}.} \bibinfo{year}{2021}\natexlab{}.
\newblock \showarticletitle{Voices of the unsung: {The} role of social presence and interactivity in building empathy in 360 video}.
\newblock \bibinfo{journal}{\emph{New Media \& Society}} \bibinfo{volume}{23}, \bibinfo{number}{8} (\bibinfo{date}{Aug.} \bibinfo{year}{2021}), \bibinfo{pages}{2230--2254}.
\newblock
\showISSN{1461-4448}
\urldef\tempurl%
\url{https://doi.org/10.1177/1461444821993124}
\showDOI{\tempurl}
\newblock
\shownote{Publisher: SAGE Publications}.


\bibitem[Porcheron et~al\mbox{.}(2023)]%
        {porcheron_cyclists_2023}
\bibfield{author}{\bibinfo{person}{Martin Porcheron}, \bibinfo{person}{Leigh Clark}, \bibinfo{person}{Stuart~Alan Nicholson}, {and} \bibinfo{person}{Matt Jones}.} \bibinfo{year}{2023}\natexlab{}.
\newblock \showarticletitle{Cyclists’ {Use} of {Technology} {While} on {Their} {Bike}}. In \bibinfo{booktitle}{\emph{Proceedings of the 2023 {CHI} {Conference} on {Human} {Factors} in {Computing} {Systems}}}. \bibinfo{publisher}{ACM}, \bibinfo{address}{Hamburg Germany}, \bibinfo{pages}{1--15}.
\newblock
\showISBNx{978-1-4503-9421-5}
\urldef\tempurl%
\url{https://doi.org/10.1145/3544548.3580971}
\showDOI{\tempurl}


\bibitem[Ptolemaeus(1845)]%
        {ptolemaeus_geographia_1845}
\bibfield{author}{\bibinfo{person}{Claudius Ptolemaeus}.} \bibinfo{year}{1845}\natexlab{}.
\newblock \bibinfo{booktitle}{\emph{Geographia}}. Vol.~\bibinfo{volume}{3}.
\newblock \bibinfo{publisher}{Tauchnitz}.
\newblock


\bibitem[Salvador et~al\mbox{.}(1999)]%
        {salvador_design_1999}
\bibfield{author}{\bibinfo{person}{Tony Salvador}, \bibinfo{person}{Genevieve Bell}, {and} \bibinfo{person}{Ken Anderson}.} \bibinfo{year}{1999}\natexlab{}.
\newblock \showarticletitle{Design {Ethnography}}.
\newblock \bibinfo{journal}{\emph{Design Management Journal (Former Series)}} \bibinfo{volume}{10}, \bibinfo{number}{4} (\bibinfo{year}{1999}), \bibinfo{pages}{35--41}.
\newblock
\showISSN{1948-7169}
\urldef\tempurl%
\url{https://doi.org/10.1111/j.1948-7169.1999.tb00274.x}
\showDOI{\tempurl}
\newblock
\shownote{\_eprint: https://onlinelibrary.wiley.com/doi/pdf/10.1111/j.1948-7169.1999.tb00274.x}.


\bibitem[Sarkar et~al\mbox{.}(2022)]%
        {sarkar_360-degree_2022}
\bibfield{author}{\bibinfo{person}{Ayush Sarkar}, \bibinfo{person}{Anh Nguyen}, \bibinfo{person}{Zhisheng Yan}, {and} \bibinfo{person}{Klara Nahrstedt}.} \bibinfo{year}{2022}\natexlab{}.
\newblock \showarticletitle{A 360-{Degree} {Video} {Analytics} {Service} for {In}-{Classroom} {Firefighter} {Training}}. In \bibinfo{booktitle}{\emph{2022 {Workshop} on {Cyber} {Physical} {Systems} for {Emergency} {Response} ({CPS}-{ER})}}. \bibinfo{pages}{13--18}.
\newblock
\urldef\tempurl%
\url{https://doi.org/10.1109/CPS-ER56134.2022.00009}
\showDOI{\tempurl}


\bibitem[Schlosser and Matthews(2022)]%
        {schlosser_designing_2022}
\bibfield{author}{\bibinfo{person}{Paul Schlosser} {and} \bibinfo{person}{Ben Matthews}.} \bibinfo{year}{2022}\natexlab{}.
\newblock \showarticletitle{Designing for {Inaccessible} {Emergency} {Medical} {Service} {Contexts}: {Development} and {Evaluation} of the {Contextual} {Secondary} {Video} {Toolkit}}. In \bibinfo{booktitle}{\emph{Proceedings of the 2022 {CHI} {Conference} on {Human} {Factors} in {Computing} {Systems}}} \emph{(\bibinfo{series}{{CHI} '22})}. \bibinfo{publisher}{Association for Computing Machinery}, \bibinfo{address}{New York, NY, USA}, \bibinfo{pages}{1--17}.
\newblock
\showISBNx{978-1-4503-9157-3}
\urldef\tempurl%
\url{https://doi.org/10.1145/3491102.3517538}
\showDOI{\tempurl}


\bibitem[Shin(2018)]%
        {shin_empathy_2018}
\bibfield{author}{\bibinfo{person}{Donghee Shin}.} \bibinfo{year}{2018}\natexlab{}.
\newblock \showarticletitle{Empathy and embodied experience in virtual environment: {To} what extent can virtual reality stimulate empathy and embodied experience?}
\newblock \bibinfo{journal}{\emph{Computers in Human Behavior}}  \bibinfo{volume}{78} (\bibinfo{date}{Jan.} \bibinfo{year}{2018}), \bibinfo{pages}{64--73}.
\newblock
\showISSN{0747-5632}
\urldef\tempurl%
\url{https://doi.org/10.1016/j.chb.2017.09.012}
\showDOI{\tempurl}


\bibitem[Sitompul and Wallmyr(2019)]%
        {lamas_analyzing_2019}
\bibfield{author}{\bibinfo{person}{Taufik~Akbar Sitompul} {and} \bibinfo{person}{Markus Wallmyr}.} \bibinfo{year}{2019}\natexlab{}.
\newblock \showarticletitle{Analyzing {Online} {Videos}: {A} {Complement} to {Field} {Studies} in {Remote} {Locations}}.
\newblock In \bibinfo{booktitle}{\emph{Human-{Computer} {Interaction} – {INTERACT} 2019}}, \bibfield{editor}{\bibinfo{person}{David Lamas}, \bibinfo{person}{Fernando Loizides}, \bibinfo{person}{Lennart Nacke}, \bibinfo{person}{Helen Petrie}, \bibinfo{person}{Marco Winckler}, {and} \bibinfo{person}{Panayiotis Zaphiris}} (Eds.). Vol.~\bibinfo{volume}{11748}. \bibinfo{publisher}{Springer International Publishing}, \bibinfo{address}{Cham}, \bibinfo{pages}{371--389}.
\newblock
\showISBNx{978-3-030-29386-4 978-3-030-29387-1}
\urldef\tempurl%
\url{https://doi.org/10.1007/978-3-030-29387-1_21}
\showDOI{\tempurl}
\newblock
\shownote{Series Title: Lecture Notes in Computer Science}.


\bibitem[Snelson and Hsu(2020)]%
        {snelson_educational_2020}
\bibfield{author}{\bibinfo{person}{Chareen Snelson} {and} \bibinfo{person}{Yu-Chang Hsu}.} \bibinfo{year}{2020}\natexlab{}.
\newblock \showarticletitle{Educational 360-{Degree} {Videos} in {Virtual} {Reality}: a {Scoping} {Review} of the {Emerging} {Research}}.
\newblock \bibinfo{journal}{\emph{TechTrends}} \bibinfo{volume}{64}, \bibinfo{number}{3} (\bibinfo{date}{May} \bibinfo{year}{2020}), \bibinfo{pages}{404--412}.
\newblock
\showISSN{1559-7075}
\urldef\tempurl%
\url{https://doi.org/10.1007/s11528-019-00474-3}
\showDOI{\tempurl}
\newblock
\shownote{Company: Springer Distributor: Springer Institution: Springer Label: Springer Number: 3 Publisher: Springer US}.


\bibitem[Star and Griesemer(1989)]%
        {star_institutional_1989}
\bibfield{author}{\bibinfo{person}{Susan~Leigh Star} {and} \bibinfo{person}{James~R. Griesemer}.} \bibinfo{year}{1989}\natexlab{}.
\newblock \showarticletitle{Institutional {Ecology}, '{Translations}' and {Boundary} {Objects}: {Amateurs} and {Professionals} in {Berkeley}'s {Museum} of {Vertebrate} {Zoology}, 1907-39}.
\newblock \bibinfo{journal}{\emph{Social Studies of Science}} \bibinfo{volume}{19}, \bibinfo{number}{3} (\bibinfo{year}{1989}), \bibinfo{pages}{387--420}.
\newblock
\showISSN{0306-3127}
\urldef\tempurl%
\url{https://www.jstor.org/stable/285080}
\showURL{%
\tempurl}
\newblock
\shownote{Publisher: Sage Publications, Ltd.}.


\bibitem[Tancik et~al\mbox{.}(2020)]%
        {tancik_stegastamp_2020}
\bibfield{author}{\bibinfo{person}{Matthew Tancik}, \bibinfo{person}{Ben Mildenhall}, {and} \bibinfo{person}{Ren Ng}.} \bibinfo{year}{2020}\natexlab{}.
\newblock \bibinfo{title}{{StegaStamp}: {Invisible} {Hyperlinks} in {Physical} {Photographs}}.
\newblock
\newblock
\urldef\tempurl%
\url{https://doi.org/10.48550/arXiv.1904.05343}
\showDOI{\tempurl}
\newblock
\shownote{arXiv:1904.05343 [cs]}.


\bibitem[Tojo et~al\mbox{.}(2021)]%
        {tojo_how_2021}
\bibfield{author}{\bibinfo{person}{Naoya Tojo}, \bibinfo{person}{Tomoko Oto}, {and} \bibinfo{person}{Sumaru Niida}.} \bibinfo{year}{2021}\natexlab{}.
\newblock \showarticletitle{How {Ethnographic} {Practices} {Are} {Reconfigured} with 360-degree {Cameras}}. \bibinfo{pages}{115--122}.
\newblock
\showISBNx{978-989-758-538-8}
\urldef\tempurl%
\url{https://doi.org/10.5220/0010639000003060}
\showDOI{\tempurl}


\bibitem[Vannini(2020)]%
        {vannini_routledge_2020}
\bibfield{author}{\bibinfo{person}{Phillip Vannini}.} \bibinfo{year}{2020}\natexlab{}.
\newblock \bibinfo{booktitle}{\emph{The {Routledge} {International} {Handbook} of ethnographic film and video}}.
\newblock \bibinfo{publisher}{Routledge}.
\newblock
\showISBNx{978-0-429-19699-7}


\bibitem[Vatanen et~al\mbox{.}(2022)]%
        {vatanen_experiences_2022}
\bibfield{author}{\bibinfo{person}{Anna Vatanen}, \bibinfo{person}{Heidi Spets}, \bibinfo{person}{Maarit Siromaa}, \bibinfo{person}{Mirka Rauniomaa}, {and} \bibinfo{person}{Tiina Keisanen}.} \bibinfo{year}{2022}\natexlab{}.
\newblock \showarticletitle{Experiences in {Collecting} 360° {Video} {Data} and {Collaborating} {Remotely} in {Virtual} {Reality}}.
\newblock \bibinfo{journal}{\emph{QuiViRR: Qualitative Video Research Reports}}  \bibinfo{volume}{3} (\bibinfo{date}{Sept.} \bibinfo{year}{2022}), \bibinfo{pages}{a0005--a0005}.
\newblock
\showISSN{2597-2456}
\urldef\tempurl%
\url{https://doi.org/10.54337/ojs.quivirr.v3.2022.a0005}
\showDOI{\tempurl}


\bibitem[Vermast and Hürst(2023)]%
        {vermast_introducing_2023}
\bibfield{author}{\bibinfo{person}{Alissa Vermast} {and} \bibinfo{person}{Wolfgang Hürst}.} \bibinfo{year}{2023}\natexlab{}.
\newblock \showarticletitle{Introducing {3D} {Thumbnails} to {Access} 360-{Degree} {Videos} in {Virtual} {Reality}}.
\newblock \bibinfo{journal}{\emph{IEEE Transactions on Visualization and Computer Graphics}} \bibinfo{volume}{29}, \bibinfo{number}{5} (\bibinfo{date}{May} \bibinfo{year}{2023}), \bibinfo{pages}{2547--2556}.
\newblock
\showISSN{1941-0506}
\urldef\tempurl%
\url{https://doi.org/10.1109/TVCG.2023.3247462}
\showDOI{\tempurl}
\newblock
\shownote{Conference Name: IEEE Transactions on Visualization and Computer Graphics}.


\bibitem[Xiao et~al\mbox{.}(2012)]%
        {xiao_recognizing_2012}
\bibfield{author}{\bibinfo{person}{Jianxiong Xiao}, \bibinfo{person}{Krista~A. Ehinger}, \bibinfo{person}{Aude Oliva}, {and} \bibinfo{person}{Antonio Torralba}.} \bibinfo{year}{2012}\natexlab{}.
\newblock \showarticletitle{Recognizing scene viewpoint using panoramic place representation}. In \bibinfo{booktitle}{\emph{2012 {IEEE} {Conference} on {Computer} {Vision} and {Pattern} {Recognition}}}. \bibinfo{pages}{2695--2702}.
\newblock
\urldef\tempurl%
\url{https://doi.org/10.1109/CVPR.2012.6247991}
\showDOI{\tempurl}
\newblock
\shownote{ISSN: 1063-6919}.


\bibitem[Ylirisku and Buur(2007)]%
        {ylirisku_designing_2007}
\bibfield{author}{\bibinfo{person}{Salu~Pekka Ylirisku} {and} \bibinfo{person}{Jacob Buur}.} \bibinfo{year}{2007}\natexlab{}.
\newblock \bibinfo{booktitle}{\emph{Designing with {Video}: {Focusing} the user-centred design process}}.
\newblock \bibinfo{publisher}{Springer Science \& Business Media}.
\newblock
\showISBNx{978-1-84628-961-3}
\urldef\tempurl%
\url{https://link.springer.com/book/10.1007/978-1-84628-961-3}
\showURL{%
\tempurl}


\bibitem[Yoshida({[n.\,d.]})]%
        {yoshida_how_nodate}
\bibfield{author}{\bibinfo{person}{Takehiko Yoshida}.} \bibinfo{year}{[n.\,d.]}\natexlab{}.
\newblock \bibinfo{title}{How to use the {PazuCraft}}.
\newblock
\newblock
\urldef\tempurl%
\url{https://www.chihayafuru.jp/pazucraft/howtouse_en.html}
\showURL{%
\tempurl}


\bibitem[Zigelbaum et~al\mbox{.}(2007)]%
        {zigelbaum_tangible_2007}
\bibfield{author}{\bibinfo{person}{Jamie Zigelbaum}, \bibinfo{person}{Michael~S. Horn}, \bibinfo{person}{Orit Shaer}, {and} \bibinfo{person}{Robert J.~K. Jacob}.} \bibinfo{year}{2007}\natexlab{}.
\newblock \showarticletitle{The tangible video editor: collaborative video editing with active tokens}. In \bibinfo{booktitle}{\emph{Proceedings of the 1st international conference on {Tangible} and embedded interaction}} \emph{(\bibinfo{series}{{TEI} '07})}. \bibinfo{publisher}{Association for Computing Machinery}, \bibinfo{address}{New York, NY, USA}, \bibinfo{pages}{43--46}.
\newblock
\showISBNx{978-1-59593-619-6}
\urldef\tempurl%
\url{https://doi.org/10.1145/1226969.1226978}
\showDOI{\tempurl}


\end{thebibliography}
\appendix
\clearpage
\onecolumn
\section{Example 360° Videos}\label{apx:example-videos}

Table~\ref{tab:example-videos} contains a list of the example 360° videos provided to participants during the study. These videos were selected based on cases used in previous literature~\cite{meijer_sphere_2024,neubauer_experiencing_2017}. Participants were free to search for or request additional videos, although only P5 briefly attempted to.
\begin{table}[H]
\begin{tabular}{lll}
Video Name & Context & URL \\ \hline
Ayala Malls Cloverleaf Bike Parking 360° VR & Cycling & \url{https://youtu.be/p25FJjkWvo8} \\
Cycling to Gare du Nord Paris & Cycling & \url{https://youtu.be/5OMXDKevBeY} \\
Astrobee robots in 360° | Cosmic Kiss & Space & \url{https://youtu.be/ZfFssKBiOn8} \\
Space science in 360° | Cosmic Kiss & Space & \url{https://youtu.be/Hrg4yxhH0OM} \\
Firefighter Training - Overview & Firefighting & \url{https://youtu.be/BLx6rLj2Ziw} \\
360° cockpit view | SWISS Airbus A320 | Geneva – Zurich & Flying & \url{https://youtu.be/HEEIzZ7UjRg} \\
How To Land An Airplane | 360° Interactive Cockpit & Flying & \url{https://youtu.be/4Vb22o1NsEw} \\
BrightFarms Virtual Reality Greenhouse Tour & Gardening & \url{https://youtu.be/bcKm3yxWOQI} \\

\end{tabular}
\caption{The example videos provided to participants during the evaluation of Tangi and the artifacts it produces.}
\label{tab:example-videos}
\end{table}

\newpage
\section{Example Artifacts}\label{apx:example-artifact}
\begin{figure}[h]
    \centering
    \includegraphics[width=0.7\linewidth]{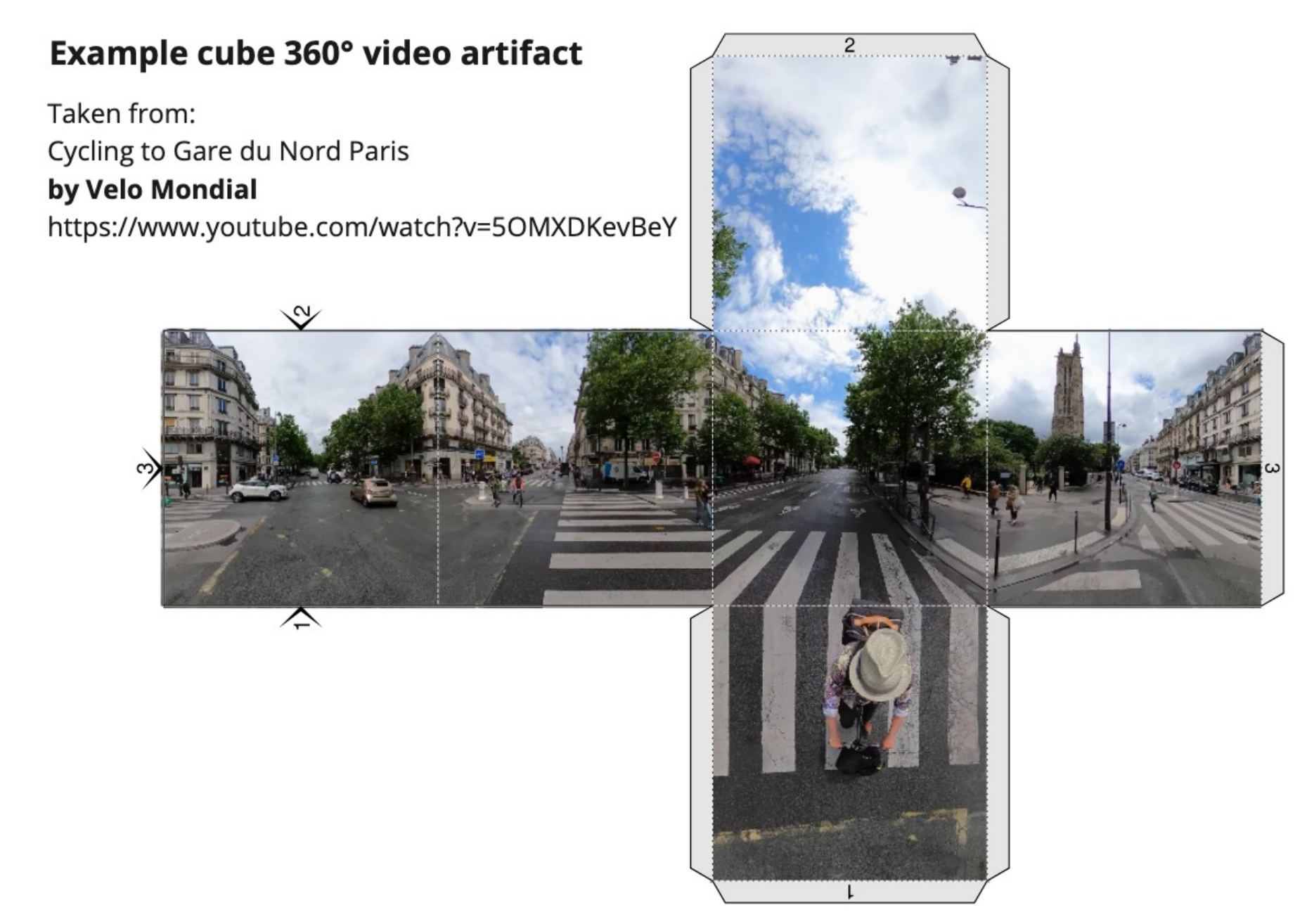}
    \caption{An example of the cube artifact generated by Tangi.}
    \Description{a 360° photo of the streets of Paris that is projected onto a flat net of a cube.}
    \label{fig:example-cube}
\end{figure}
\begin{figure}[h]
    \centering
    \includegraphics[width=0.7\linewidth]{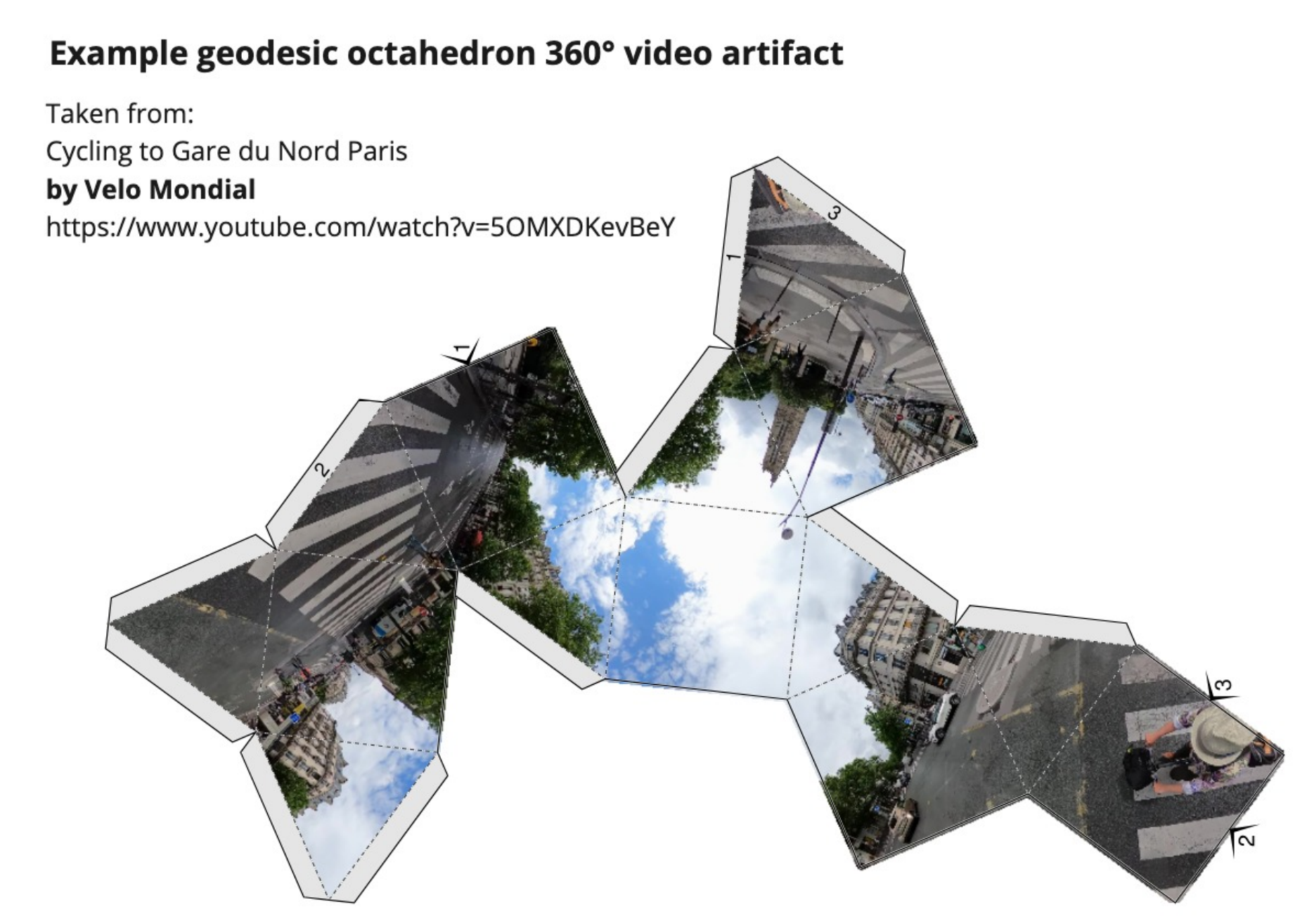}
    \caption{An example of the geodesic octahedron artifact generated by Tangi.}
    \Description{a 360° photo of the streets of Paris that is projected onto a flat net of a geodesic octahedron.}
    \label{fig:example-poly}
\end{figure}





\end{document}